\documentclass[final]{pasj00}

\begin{document}
\SetRunningHead{S.~Ozaki}{NLR Structure of NGC~1068 }
\Received{0000/00/00}
\Accepted{0000/00/00}

\title{Kinematic and Excitation Structure of\\ the NGC~1068 Narrow-Line Region}

\author{Shinobu \textsc{Ozaki}}
\affil{Okayama Astrophysical Observatory, National Astronomical Observatory of Japan,\\
Kamogata, Asakuchi, Okayama 719-0232, Japan}
\email{ozaki@oao.nao.ac.jp}


\maketitle

\begin{abstract}
We investigated the kinematic and excitation structure of the NGC~1068 narrow-line region (NLR).
We obtained profiles of several emission lines, [O\emissiontype{III}]$\lambda$5007, H$\beta$, [O\emissiontype{I}]$\lambda$6300 and [Fe\emissiontype{VII}]$\lambda$6087 at high-velocity resolution (R $\sim$ 7500 - 11000), and confirmed that they showed different profiles.
These profiles are useful for understanding the NLR structure, as they cover a wide ionization potential range.
By comparing the results with a photoionization model, we found that 
1) blueshifted components at the center are very dense, 
2) those in the northeast region have slightly lower densities than those in the center, 
and 3) ionization parameters of the blueshifted components increase with increasing velocity with respect to the systemic velocity.
We investigated the NLR structure in NGC~1068 based on these results.
We show that both the observed velocity dependence of the ionization parameter and the gradually increasing velocity field can be reproduced by varying the ionizing continuum attenuation, assuming a hollowed biconical geometry and varying the column densities of outflowing clouds.
\end{abstract}

\section{Introduction} \label{sec:intro}

Narrow-line region (NLR) structure is one of the most interesting aspects of Seyfert galaxies and, consequently, has been long studied.
One tool in such investigations is an emission-line profile.
A blueward asymmetry in narrow-line profiles is frequently observed in Seyfert galaxies, and this has led to two traditional interpretations: an outflow model including extinction between line-emitting clouds, and
an inflow model in which extinction originates in individual clouds 
(e.g., \cite{whittle1985a,derobertis1990,veilleux1991a}).
In some Seyfert galaxies, emission lines with higher critical electron density and/or higher ionization potential have lower central velocities and/or wider line widths.
These features have been interpreted as implying that, in the NLR, velocity decreases with increasing distance from the nucleus, and that the regions closer to the nucleus are denser and in a higher excitation state (e.g., \cite{whittle1985b,derobertis1990,veilleux1991b}).

NGC~1068 is one of the most famous Seyfert galaxies and has been studied extensively.
The distance to NGC~1068 is 14.4 Mpc (\timeform{1''} corresponds to 70 pc.) and its systemic velocity is 1148 km~s$^{-1}$ \citep{bland}.
Although NGC~1068 has been considered a prototypical Seyfert 2 galaxy, its forbidden lines are much wider than those of typical Seyfert 2 galaxies, and the emission-line profiles show multiple components \citep{pelat,alloin,meaburn1986}.
\citet{walker} suggested outflow in the NLR of NGC~1068, and \citet{cecil1990} and \citet{arribas} argued that the velocity field of the NLR can be explained as a biconical outflow, based on tri-dimensional spectroscopic studies.

The excellent spatial resolution of the {\it Hubble Space Telescope} ({\it HST}) revealed that the striking velocity structure of the NLR gradually increased with radius (Crenshaw \& Kraemer 2000, hereafter \authorcite{ck_kinematics2000}).
It must be noted that such a velocity field differs from the classical perspective, which states that higher velocity components lie closer to the nucleus.
The velocity field was shown to be reproducible with a hollowed biconical outflow model (\cite{das2006,cecil2002}; \authorcite{ck_kinematics2000}), but the radial dependence of the velocity in the model was included without a physical rationale.
\citet{das2007} and \citet{everett} attempted to explain the radial dependence of the velocity, but no successful explanation has been suggested to date.

\begin{figure*}
	\begin{center}
		\FigureFile(110mm,114mm){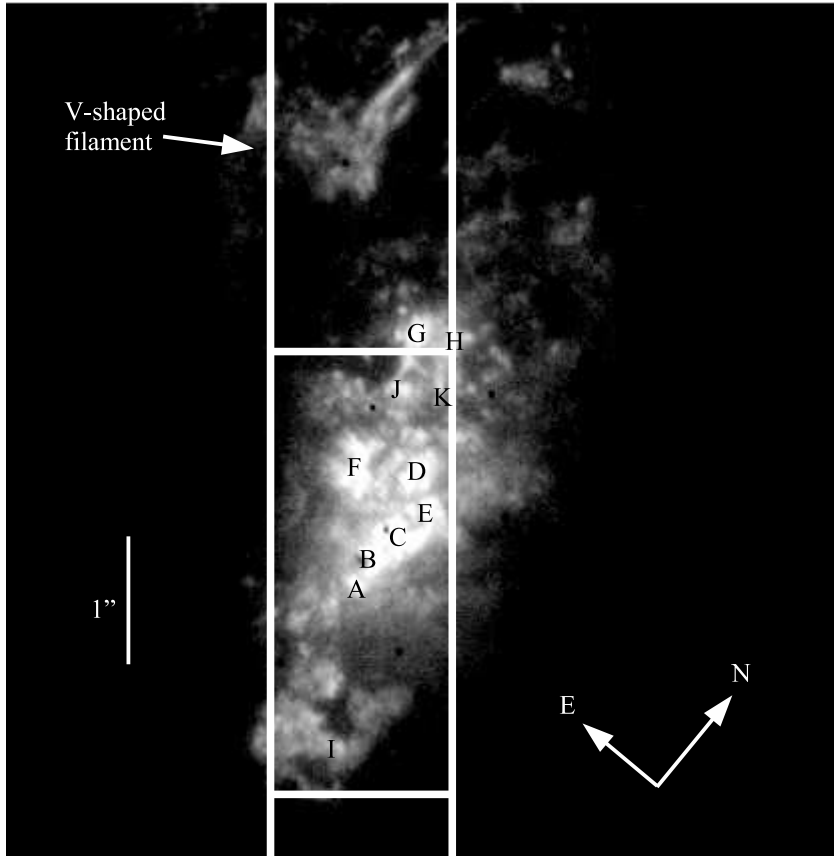}
	\end{center}
	\caption{FOC image of [O\emissiontype{III}]$\lambda\lambda 4959,5007$ (\cite{macchetto};
Data were downloaded from the Multimission Archive at the Space Telescope Science Institute\footnotemark.). Two white vertical lines indicate the \timeform{1.2''} slit width aligned for the typical PA=$40^{\circ}$. 
 Extraction bins (\timeform{3.4''}) are also shown. Letters from A to K indicate the cloud locations reported by \citet{evans} and \citet{cecil2002}.}
	\label{fig:clouds}
\end{figure*}

\begin{table*}
	\begin{center}
	\caption{Observation log}
	\label{tab:obslog}
		\begin{tabular}{llllll}
			\hline
			\rule[0pt]{0pt}{10pt}Date & Object & Wavelength range & Included lines& Exposure time & PA\\[2pt] 
			\hline
			\rule[0pt]{0pt}{12pt}30 Nov 2005 & NGC~1068  & 4750 - 5190 {\AA} & H$\beta$, [O\emissiontype{III}] & 1800 sec & 37 - 41$^{\circ}$\\
			           & NGC~1068   & 4750 - 5190 {\AA} & H$\beta$, [O\emissiontype{III}] & 1800 sec & 38 - 39$^{\circ}$\\
			           & HD~30739  & 4750 - 5190 {\AA} & & 100 sec & \\
			24 Jan 2006 & NGC~1068  & 6020 - 6440 {\AA}  & [Fe\emissiontype{VII}], [O\emissiontype{I}] & 1800 sec & 36 - 45$^{\circ}$\\
				      & NGC~1068  & 6020 - 6440 {\AA}  & [Fe\emissiontype{VII}], [O\emissiontype{I}] & 1800 sec & 32 - 39$^{\circ}$\\
					  & HD~74280 & 6020 - 6440 {\AA} & & 150 sec   &  \\
			\hline
		\end{tabular}
	\end{center}
\end{table*}

The observed emission-line intensity ratios of the NGC~1068 NLR can be reproduced by photoionization models
(\cite{groves,kc2000II}; Kraemer \& Crenshaw 2000b, hereafter \authorcite{kc2000III}; \cite{alexander,lutz}).
\authorcite{kc2000III} decomposed the profiles into two kinematic components and investigated the physical condition of each component.
They reported that, while blueshifted clouds suffer from direct ionizing continuum, redshifted clouds are irradiated by an attenuated ionizing continuum.
However, the coupling of excitation structure to kinematics has not been studied in detail.

In this study, we investigated the kinematic and excitation structure of the NGC~1068 NLR in detail.
We performed medium-resolution spectroscopy for NGC~1068 and obtained high velocity resolution [O\emissiontype{III}]$\lambda$5007, H$\beta$, [O\emissiontype{I}]$\lambda$6300 and [Fe\emissiontype{VII}]$\lambda$6087 emission-line profiles.
As the ionization potentials of these emission lines encompass a wide range, 0 eV to 100 eV, we can research the physical condition of the NLR over a wide range of excitation states.

The outline of this paper is as follows:
We describe observations and data reduction in section 2, 
and summarize features of the obtained profiles and velocity dependences of the emission-line ratios in section 3.
In section 4, we compare the observational results to a photoionization model.
In section 5, we discuss the relationship between ionization potentials and the central velocities of emission lines, and then investigate the kinematic and excitation structure of the NGC~1068 NLR.

\footnotetext{http://archive.stsci.edu/}

\section{Observations and data reduction}

\subsection{Observations}
We performed medium-resolution spectroscopy of the nuclear region of NGC~1068 with a new optical spectrograph \citep{malls} mounted 
at the Nasmyth focus (F/12) of the 2-m NAYUTA telescope at Nishi-Harima Astronomical Observatory on 30 November 2005 and 24 January 2006.
The observation log is summarized in table~\ref{tab:obslog}.
Weather conditions were clear and stable for the first night,
and almost clear with occasional clouds for the second.
The seeing was approximately 2 arcsec for each day.
On 24 January 2006, however, the spatial resolution degraded significantly because of seeing variations or telescope tracking errors during the exposures.

Slit width and length were \timeform{1.2''} and \timeform{5'}, respectively.
A grating with 1800 grooves~mm$^{-1}$ was used.
This combination of slit and grating provided spectral resolutions of 7500 at 5000 {\AA}, and 11000 at 6220 {\AA}, corresponding to velocity resolutions of 40 km~s$^{-1}$ and 27 km~s$^{-1}$, respectively.

We used the NGC~1068 nucleus as a marker to guide the telescope.
Unfortunately, the image rotator did not function well; hence, the slit swept out some area during each exposure due to image rotation.
The position angle (PA) range for each exposure is listed in table~\ref{tab:obslog}.
After each exposure, we tried to bring the PA back to the initial position.
We were successful on 30 November 2005 but not on 24 January 2006.
The maximum difference in PA was $13^{\circ}$.
The NE and SW extraction windows were located at \timeform{3.4''} from the center; $13^{\circ}$ corresponds to \timeform{0.77''} for the NE and SW extraction windows.
This difference does not affect the results significantly, as the seeing size was approximately \timeform{2''}.
The typical PA was 40$^{\circ}$, which is approximately aligned to a radio jet axis (PA=32$^{\circ}$; \cite{wilson}).
Figure~\ref{fig:clouds} shows the average slit position.

We also observed photometric standard stars HD~30739 and HD~74280 for flux calibration on the same nights.

\subsection{Data reduction}
\label{sec:dataanalysis}
We reduced the data with IRAF\footnote{IRAF is distributed by the National Optical Astronomy Observatories, which are operated by the Association of Universities for Research in Astronomy, Inc., under cooperative agreement with the National Science Foundation.}.
We performed bias subtraction by referring to overscan regions, and then made flat-field corrections using flat-field frames made by combining three flat frames obtained after each set of exposures.
We performed wavelength calibrations using comparison frames from a hollow cathode lamp (FeNeAr).
The root mean square error for the wavelength calibration was less than 0.1 {\AA}.
We checked the systematic error of the wavelength calibration with a night-sky line of [O\emissiontype{I}]$\lambda$6300.3, and measured the centers of the line as 6300.1 {\AA} and 6300.2 {\AA} for each frame.
We concluded that wavelength calibration accuracy was better than 0.2 {\AA}.
After background subtraction, we ran a flux calibration and rejected cosmic ray events.
Finally, we applied the correction for heliocentric velocity.

\begin{figure}
	\begin{center}
		\FigureFile(80mm,60mm){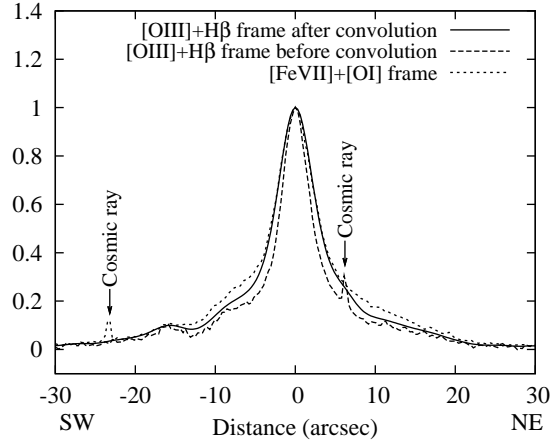}
	\end{center}
	\caption{Comparison of continuum radial profiles normalized by peak intensity.}
	\label{fig:radial_profile_continuum}
\end{figure}

Continuum radial profiles around the center of the [O\emissiontype{III}]+H$\beta$ frame and the [Fe\emissiontype{VII}]+[O\emissiontype{I}] frame were different.
Comparing g-, r- and i-band images of NGC~1068 downloaded from the Isaac Newton Group Archive\footnote{http://casu.ast.cam.ac.uk/casuadc/archives/ingarch}, we confirmed the absence of any considerable variation in the continuum radial profile with wavelength.
Thus, this difference is thought to have been caused by seeing variations or telescope tracking errors during the exposure.
We matched the continuum radial profile in the [O\emissiontype{III}]+H$\beta$ frame to that of the [Fe\emissiontype{VII}]+[O\emissiontype{I}] frame using the Gaussian convolution along the spatial direction.
The resultant spatial resolution after the convolution was \timeform{3.4''}.
The results of this procedure are shown in figure~\ref{fig:radial_profile_continuum}.
Spectra at the continuum peak, NE and SW were extracted using a bin length of \timeform{3.4''} along the slit.

Flux at the center is thought to contaminate flux at the NE\timeform{3.4''} and SW\timeform{3.4''}, since the flux at the center is much stronger.
Thus, we estimated central flux contamination for the NE\timeform{3.4''} and SW\timeform{3.4''} and subtracted it from each spectrum,
assuming that the central spectrum was a spatially extended Gaussian function with a \timeform{3.4''} FWHM.

\begin{figure*}
	\begin{center}
		\FigureFile(160mm,65mm){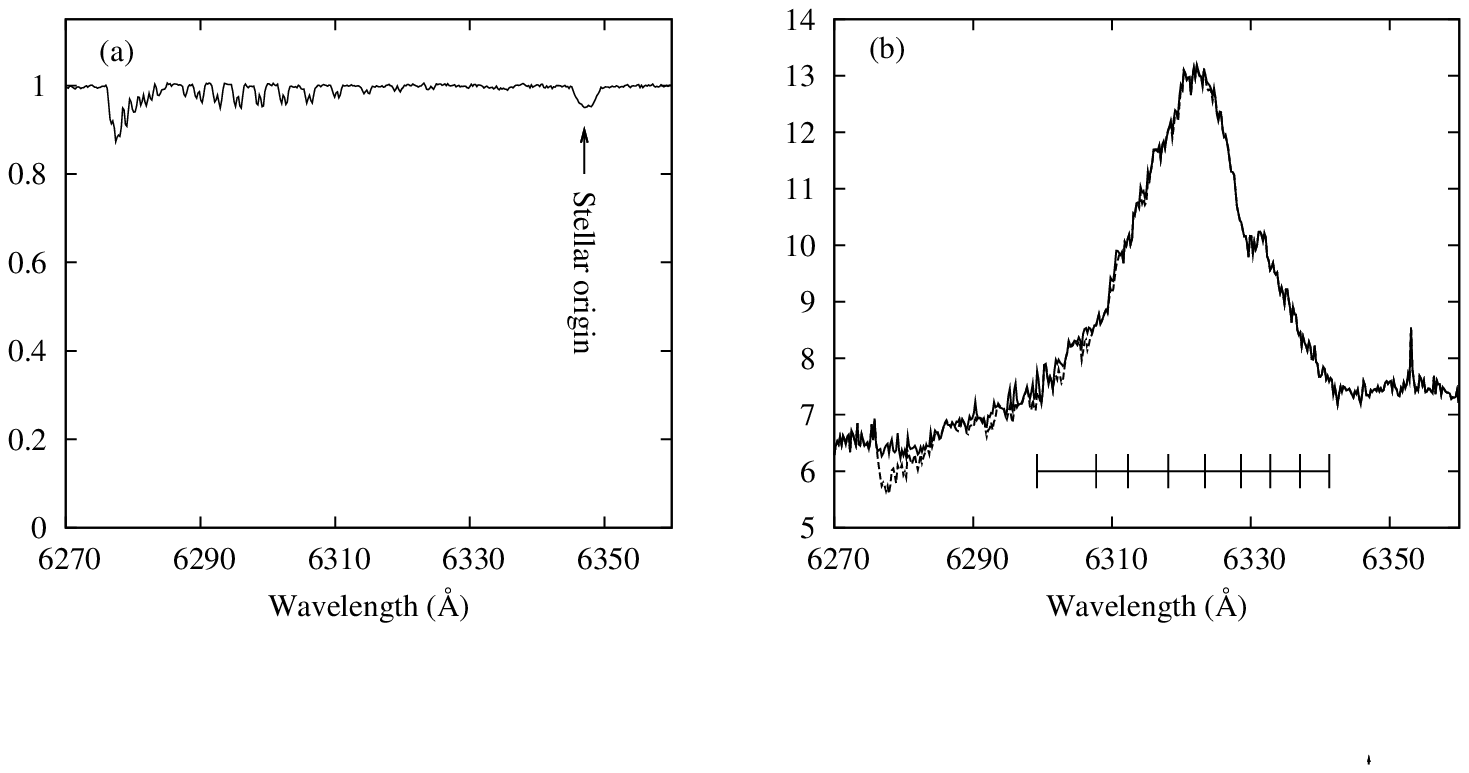}
	\end{center}
	\caption{Left panel: Normalized HD~74280 spectrum. Right panel: Dashed and solid lines show the spectra at the center around [O\emissiontype{I}]$\lambda6300$ before and after the atmospheric absorption line correction, respectively. The vertical axis represents relative strength. Flux integration ranges are also shown (subsection \ref{sec:ratio}).}
	\label{fig:atom_abs}
\end{figure*}

We applied an atmospheric absorption-line correction around [O\emissiontype{I}]$\lambda6300$.
Figure~\ref{fig:atom_abs}a shows a spectrum of HD~74280 normalized by its continuum level.
As this star's spectral type is B3V, it has few intrinsic absorption lines, so the atmospheric absorption lines are clearly distinguishable.
We applied an atmospheric absorption-line correction for the 6273 - 6327 {\AA} range by dividing the spectrum of NGC~1068 by the normalized standard star spectrum.
Figure~\ref{fig:atom_abs}b shows the spectrum of NGC~1068 before and after the correction.
We can see the negligible effect for the [O\emissiontype{I}]$\lambda6300$ profile.

\begin{table}
	\begin{center}
	\caption{Selected stars for the template stellar spectra}
	\label{tab:template}
		\begin{tabular}{ccc} \hline
			\rule[0pt]{0pt}{10pt}Spectral type & Name & [Fe/H] \\[2pt] \hline
			\rule[0pt]{0pt}{12pt}F5III & HD~171802 & 0.10 \\
			G0III & HD~039833 & 0.04 \\
			G5III & HD~027022 & 0.13 \\
			K0III & HD~130322 & 0.06 \\
			K5III & HD~164058 & -0.07 \\ \hline
		\end{tabular}
	\end{center}
\end{table}

\subsection{Stellar component subtraction}
Stellar components should be subtracted from the object spectra to obtain accurate emission-line fluxes.
The stellar component is a composite of various stellar spectral types.
However, spectral synthesis is difficult, as the spectra in this study cover only a short wavelength range.
Hence, we assumed that the stellar component in each spectral range could be represented by a single spectral type.

We selected some stellar spectra from F5III to K5III from the ELODIE stellar library \citep{elodie}, since bulge light is dominated by late-type red giant stars.
As the metallicity of the stellar component of NGC~1068 is uncertain, we selected stars with metallicities close to the solar value.
The selected stars are listed in table~\ref{tab:template}.

\begin{figure*}
	\begin{center}
		\FigureFile(160mm,125mm){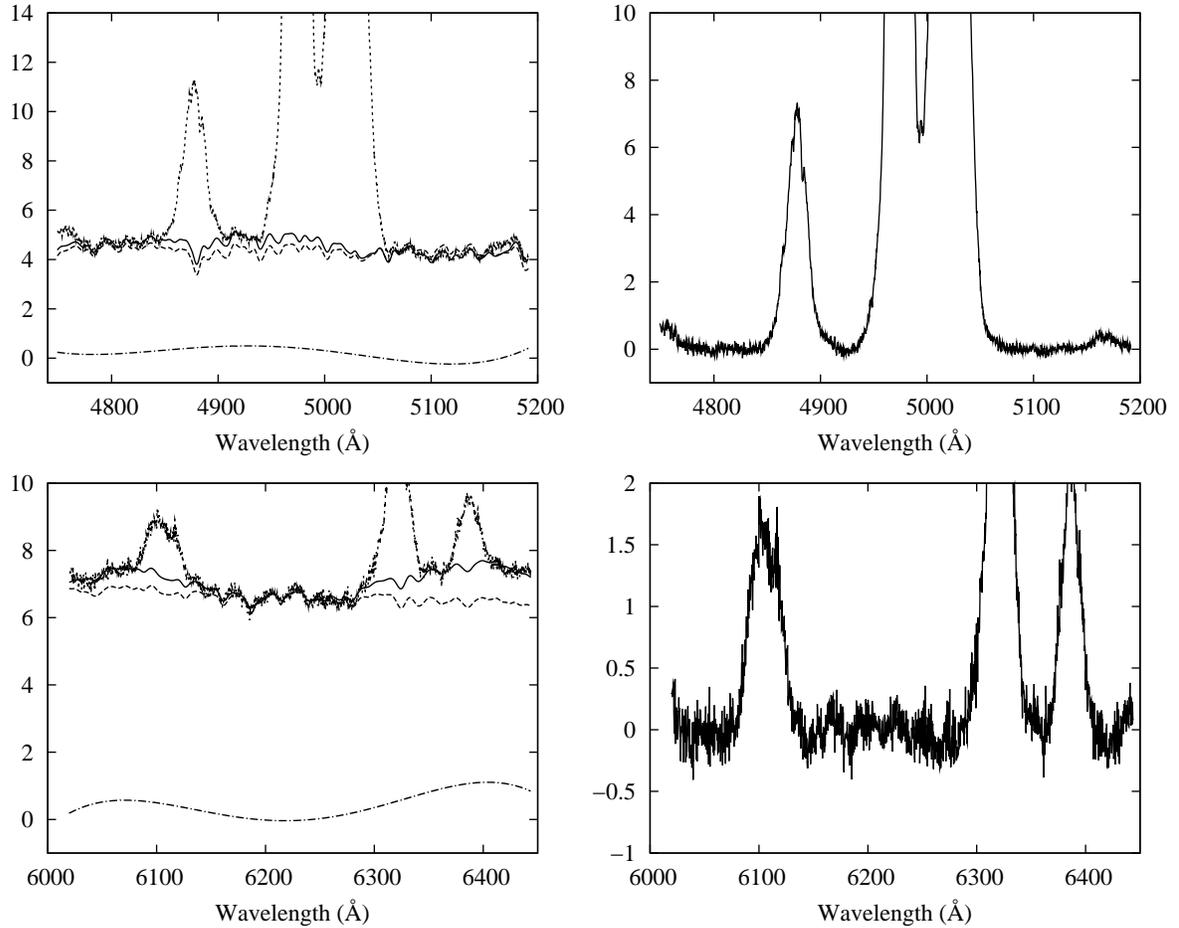}
	\end{center}
	\caption{Results of the stellar continuum subtraction at the center. Left panels show the object spectra before continuum subtraction (dotted line), and the adopted template spectrum (solid line), composed of the stellar spectrum (dashed line) and the polynomial (dot-dashed line). Right panels show the spectrum after subtraction.}
	\label{fig:continuum_sub}
\end{figure*}

We created template spectra as follows:
Spectra in the library were masked at cosmic-ray events or telluric absorption lines.
The masked regions were typically a few pixels.
We corrected the masked regions with a linear interpolation, and shifted the spectra of these stars by the recession velocity of NGC~1068. Then we convolved the spectra with velocity dispersions of 50, 100, 150, 200, 250, and 300 km~s$^{-1}$.
We created many template spectra with various combinations of spectral types and velocity dispersions.
We then scaled the template spectra by referring to line-free regions and subtracted the scaled template spectra from the object spectra.

\begin{figure*}
	\begin{center}
		\FigureFile(160mm,125mm){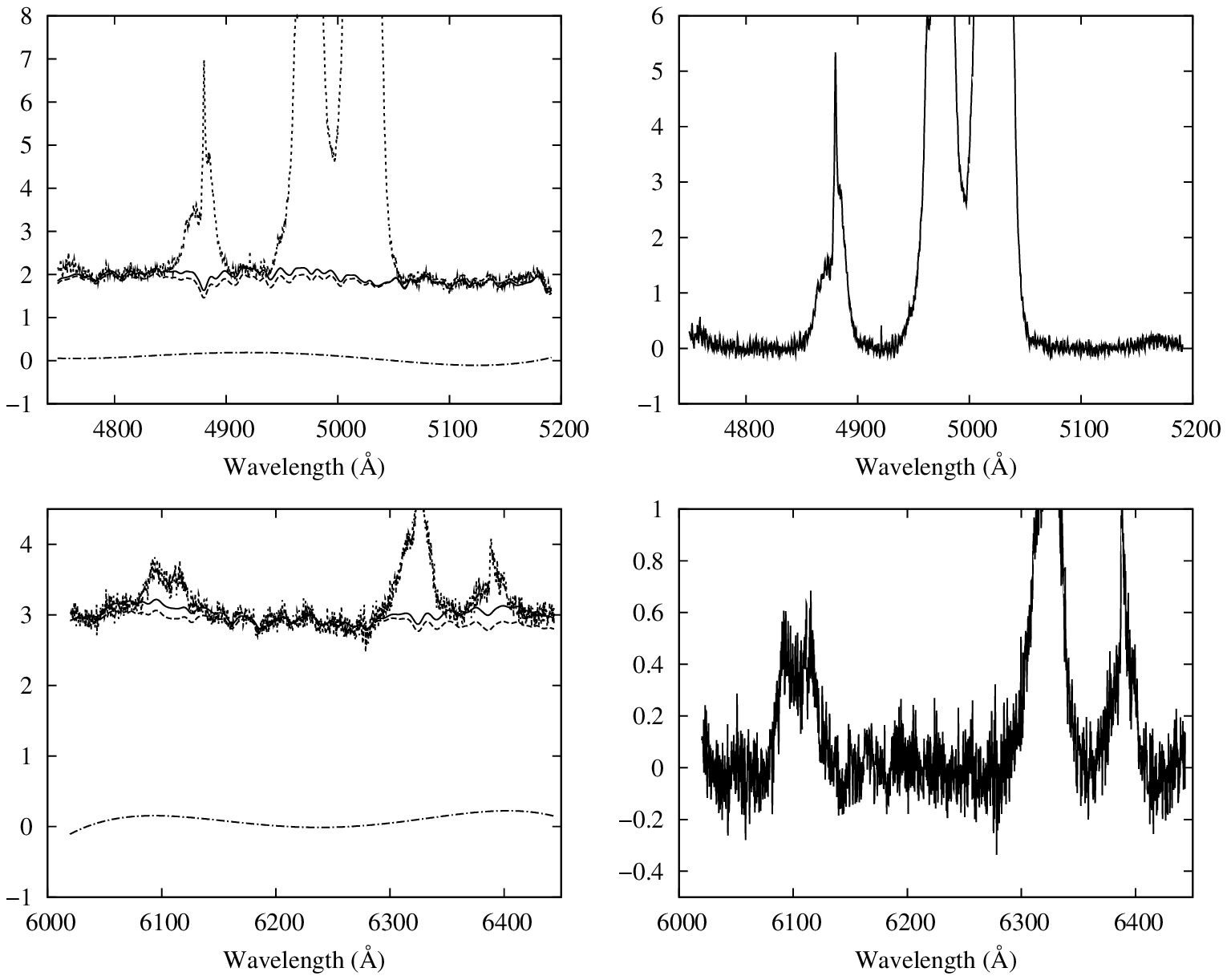}
	\end{center}
	\caption{Same as figure~\ref{fig:continuum_sub}, but for the NE\timeform{3.4"}.}
	\label{fig:continuum_sub_ne}
\end{figure*}

We compared the subtracted spectra by eye, and found that, when we adopted the G5III spectrum convolved with a 150 km~s$^{-1}$ velocity dispersion, the stellar-origin absorption lines in the object spectra were canceled out most precisely in both wavelength regions.
Therefore, we adopted this case as a best-fit template for the underlying stellar spectrum.
Using G0III or K0III with a dispersion of 100 km~s$^{-1}$ or 200 km~s$^{-1}$ resulted in no significant difference from the best-fit template.

After the template spectrum subtraction, some residuals appeared (figure~\ref{fig:continuum_sub}), 
especially around [O\emissiontype{I}]$\lambda\lambda6300,6364$.
These residuals can be explained to some extent by K5III star contamination.
We fit these residuals with a fourth-order polynomial for each spectrum and subtracted it.
The results of this continuum subtraction are shown in figure~\ref{fig:continuum_sub}.
Deviations around 4750 {\AA} and 5170 {\AA} are probably due to faint [Ar\emissiontype{IV}]$\lambda4740$ and [Fe\emissiontype{VII}]$\lambda5159$ emission lines.

\subsection{Scaling between different wavelength regions}
\label{sec:obs_scaling}
We scaled the spectra obtained on the different nights to reproduce the reddening-corrected [O\emissiontype{I}]/[O\emissiontype{III}] ratio reported by \citet{ho}.
In this procedure, reddening corrections were applied simultaneously.
\citet{ho} adopted the color excess $E(B-V)=0.54$ derived from the Balmer decrement.
\citet{koski} reported $E(B-V)=0.52$ for NGC~1068.
\authorcite{kc2000III} reported $E(B-V)= 0.13$ - $0.55$ and noted that the blueshifted components were more heavily reddened than the redshifted ones.
Thus, the extinction for the redshifted components might be overestimated in this study.

\section{Results}
\subsection{Line profiles}

\begin{figure*}
	\begin{center}
		\FigureFile(160mm,125mm){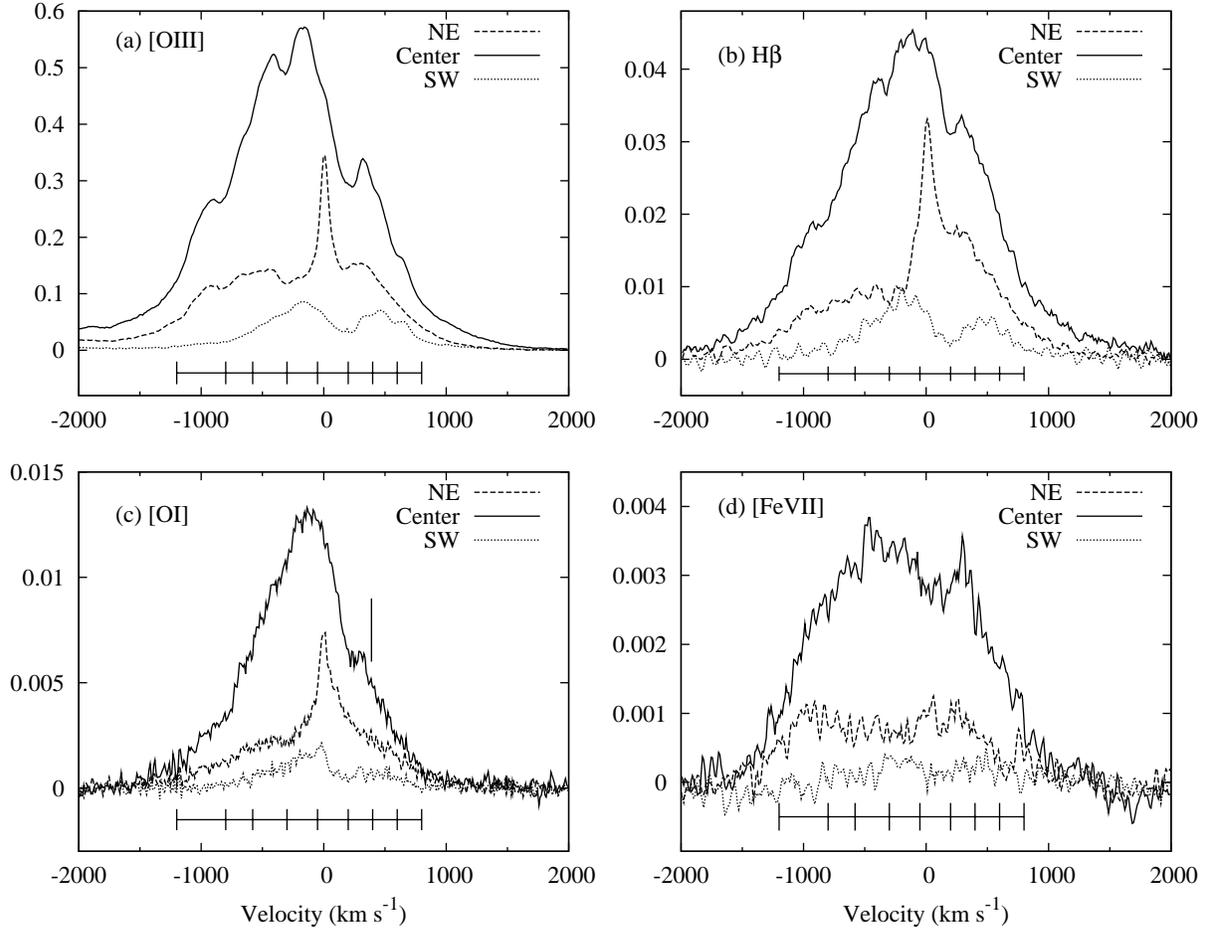}
	\end{center}
	\caption{Emission line profiles. The [Fe\emissiontype{VII}] profiles are smoothed with a 3-pixel running mean. The horizontal axis is the velocity with respect to the systemic velocity, and the vertical axis is relative intensity. Integration ranges are shown in each panel. The expected velocity of [S\emissiontype{III}]$\lambda 6312$ is shown as a vertical line in the bottom-left panel.}
	\label{fig:profile_changes}
\end{figure*}

\begin{figure}
	\begin{center}
		\FigureFile(80mm,185mm){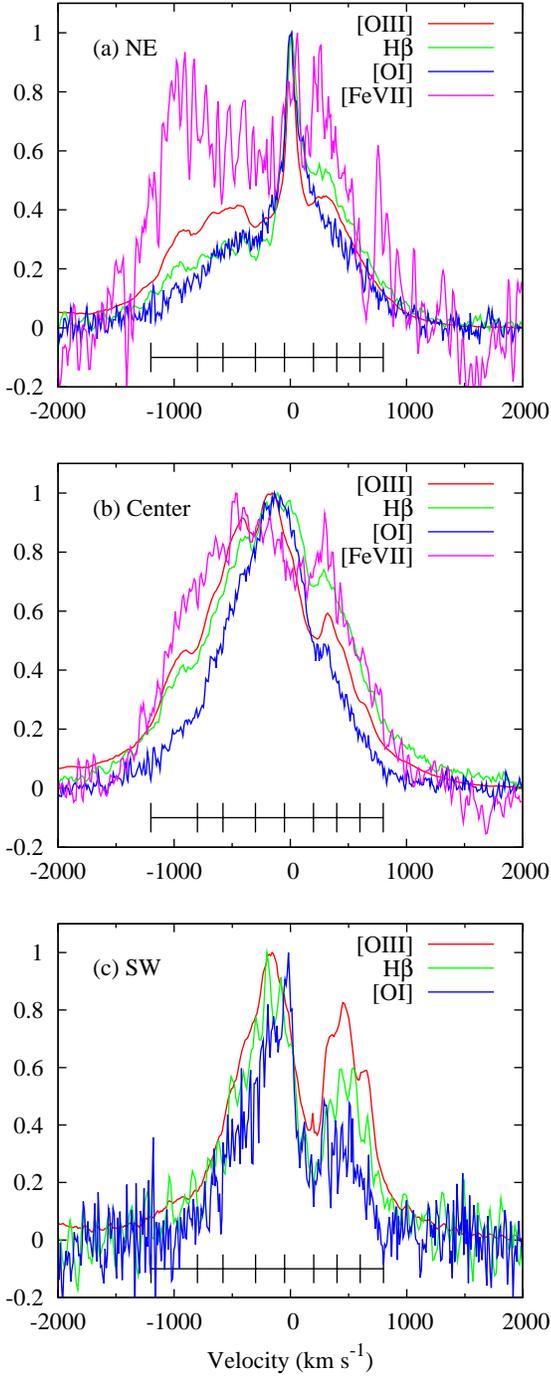}
	\end{center}
	\caption{Comparison of the line profiles at each location. Each line profile is normalized by its peak intensity. The [Fe\emissiontype{VII}] profiles are smoothed with a 3-pixel running mean. The [Fe\emissiontype{VII}] profile at the SW\timeform{3.4''} is not shown because of its low S/N ratio. Integration ranges are shown in each panel.}
	\label{fig:profile_place}
\end{figure}

NGC~1068 has been studied extensively and many kinds of observations have been performed.
However, a detailed comparison of several emission lines at high-velocity resolution, such as presented in this study, has not been performed except for [O\emissiontype{III}] and H$\beta$ \citep{alloin}.
Figure~\ref{fig:profile_changes} shows the emission-line profiles.
In figure~\ref{fig:profile_place}, we compare the profiles at each position.
The [Fe\emissiontype{VII}] profiles were smoothed with a 3-pixel running mean in figures~\ref{fig:profile_changes} and \ref{fig:profile_place}.
In figure~\ref{fig:profile_place}, the [Fe\emissiontype{VII}] profile for the SW \timeform{3.4''} is not shown because of its low signal-to-noise ratio.
In the following subsubsections, we describe the features of the line profiles at each position.

\begin{table}
	\begin{center}
	\caption{Adopted wavelength.}
	\label{tab:wavelength}
		\begin{tabular}{rc} \hline
			\rule[0pt]{0pt}{10pt}Line name & Wavelength \\[2pt] \hline
 			\rule[0pt]{0pt}{12pt}~H$\beta$ & 4861.3 \AA \\
			~[O\emissiontype{III}]$\lambda5007$ & 5006.8 \AA \\
			~[Fe\emissiontype{VII}]$\lambda6087$ & 6087.0 \AA \\
			~[O\emissiontype{I}]$\lambda6300$ & 6300.3 \AA \\[2pt] \hline
		\end{tabular}
	\end{center}
\end{table}

\begin{figure}
	\begin{center}
		\FigureFile(80mm,70mm){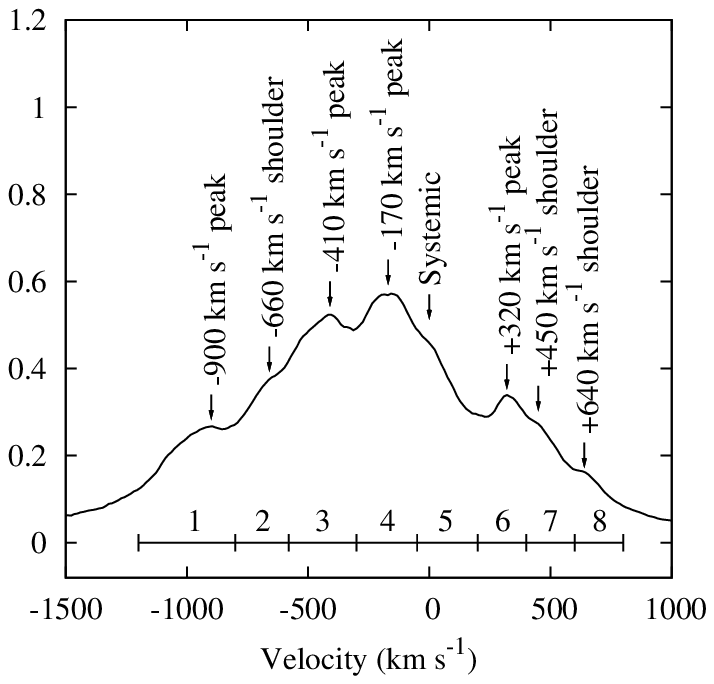}
	\end{center}
	\caption{The [OIII] profile at the center. The features referred to in the text, the integration ranges and the ID numbers of the velocity bins are shown.}
	\label{fig:explain_features}
\end{figure}

\subsubsection{Line profiles at the center}
\label{sec:CentralLineProfile}

The [O\emissiontype{III}]$\lambda$5007 profile has four peaks and many shoulders.
These features are consistent with previous high-dispersion spectroscopic studies \citep{meaburn1986,profile1998}.
The H$\beta$ profile is globally similar to the [O\emissiontype{III}] profile.

An [O\emissiontype{I}] profile peak was found at the same velocity ($\sim -170$ km~s$^{-1}$) as [O\emissiontype{III}] and H$\beta$. 
This means that the peak of each line originates in the same cloud and that the cloud is radiation-bounded because it has a partially ionized zone.

The [O\emissiontype{I}] profile displays a narrower shape than the others.
In the [O\emissiontype{I}] profile, we found no well-defined peaks around $-410$ km~s$^{-1}$ and $-900$ km~s$^{-1}$, where [O\emissiontype{III}] and H$\beta$ exhibit peaks.
However we recognized a shoulder around $-410$ km~s$^{-1}$ and a weak tail around $-900$ km~s$^{-1}$.
The +320 km~s$^{-1}$ peak of the [O\emissiontype{I}] profile is much weaker than the others.
These features suggest that [O\emissiontype{I}] flux is suppressed in the high-velocity clouds.

The +320 km~s$^{-1}$ peak of the [O\emissiontype{I}] profile may be produced by [S\emissiontype{III}]$\lambda6312$ which is separated from [O\emissiontype{I}]$\lambda6300$ by about +560 km~s$^{-1}$.
If [S\emissiontype{III}] has the same velocity as H$\beta$, [O\emissiontype{III}] and [O\emissiontype{I}], the expected [S\emissiontype{III}] peak would be located at +390 km~s$^{-1}$ with respect to the systemic velocity of [O\emissiontype{I}]. Its position is marked in figure~\ref{fig:profile_changes}c.
A displacement between the expected peak location and the +320 km~s$^{-1}$ peak is clear. No feature was found at the expected location.
The +320 km~s$^{-1}$ peak of [O\emissiontype{I}] is consistent with those of the other lines (figure~\ref{fig:profile_place}b).
Hence, we conclude that the contribution from [S\emissiontype{III}] is negligible.

The double-peaked [Fe\emissiontype{VII}] profile is quite different from the other profiles at first glance, as reported by \citet{rodriguez}.
\citet{capetti1997}, in a narrow-band imaging study with {\it HST}, showed that [Ne\emissiontype{V}] originates from the [O\emissiontype{III}]-bright NLR clouds.
The ionization potentials of Ne$^{4+}$ and Ne$^{5+}$ are 97 eV and 126 eV, respectively, similar to those of Fe$^{6+}$ (100 eV) and Fe$^{7+}$ (128 eV).
Thus, it is plausible that [Fe\emissiontype{VII}] is emitted from the [O\emissiontype{III}]-bright NLR clouds.

A detailed comparison of the [Fe\emissiontype{VII}] profile with the others reveals the following three features:
First, the blueshifted and redshifted [Fe\emissiontype{VII}] peaks correspond to the [O\emissiontype{III}] peaks at $-410$ km~s$^{-1}$ and +320 km~s$^{-1}$, respectively.
Second, no counterpart appears in the [Fe\emissiontype{VII}] profile for the $-170$ km~s$^{-1}$ peak seen in the other lines.
Third, in the wavelength region bluer than $-410$ km~s$^{-1}$, [Fe\emissiontype{VII}] is relatively stronger than the other lines.

\subsubsection{Line profiles at NE}
The [O\emissiontype{III}] profile has wide bumps in the bluer part, while the [O\emissiontype{I}] profile declines smoothly.
The [O\emissiontype{I}] profile does not exhibit a peak around +320 km~s$^{-1}$; the other lines do.
These features indicate that [O\emissiontype{I}] is suppressed in high-velocity components like the center.

A bright narrow component around the systemic velocity called a ``velocity spike'' by \citet{meaburn1986} is predominant except for the [Fe\emissiontype{VII}] profile.
This component was marginally detected in the [Fe\emissiontype{VII}] profile, 
but its relative strength was considerably lower than that of the other emission lines.
The blueshifted peak of the [Fe\emissiontype{VII}] profile is bluer at the NE\timeform{3.4''} than the center, as pointed out by \citet{rodriguez}. 

\subsubsection{Line profiles at SW}
The profiles, except [Fe\emissiontype{VII}], show similar shapes.
[O\emissiontype{III}] is relatively stronger in the redshifted part than the other lines.
This result is consistent with \authorcite{kc2000III}, who reported that, in the SW region of the ionization cone, the redshifted components show higher excitation than the blueshifted components.

The profiles, except [O\emissiontype{III}], have low signal-to-noise ratios.
Therefore, we did not use the SW data in the following sections. 

\subsection{Line intensity ratios}
\label{sec:ratio}

\begin{table}
	\begin{center}
	\caption{Velocity ranges for integration.}
	\label{tab:range}
		\begin{tabular}{ccrl} \hline
			\rule[0pt]{0pt}{10pt}ID & Feature name & \multicolumn{2}{c}{Velocity range} \\[2pt] \hline
			\rule[0pt]{0pt}{12pt}1 & $-900$ km~s$^{-1}$ peak & $-1200$ -& \hspace{-3mm}$-800$ km~s$^{-1}$ \\
			2 & $-660$ km~s$^{-1}$ shoulder & $-800$ -& \hspace{-3mm}$-580$ km~s$^{-1}$ \\
			3 & $-410$ km~s$^{-1}$ peak & $-580$ -& \hspace{-3mm}$-300$ km~s$^{-1}$ \\
			4 & $-170$ km~s$^{-1}$ peak & $-300$ -& \hspace{-3mm}$-50$ km~s$^{-1}$ \\
			5 & Systemic & $-50$ -& \hspace{-3mm}+200 km~s$^{-1}$ \\
			6 & +320 km~s$^{-1}$ peak & +200 -& \hspace{-3mm}+400 km~s$^{-1}$ \\
			7 & +450 km~s$^{-1}$ shoulder & +400 -& \hspace{-3mm}+600 km~s$^{-1}$ \\
			8 & +640 km~s$^{-1}$ shoulder & +600 -& \hspace{-3mm}+800 km~s$^{-1}$ \\ \hline
		\end{tabular}
	\end{center}
\end{table}

\begin{figure}
	\begin{center}
		\FigureFile(80mm,185mm){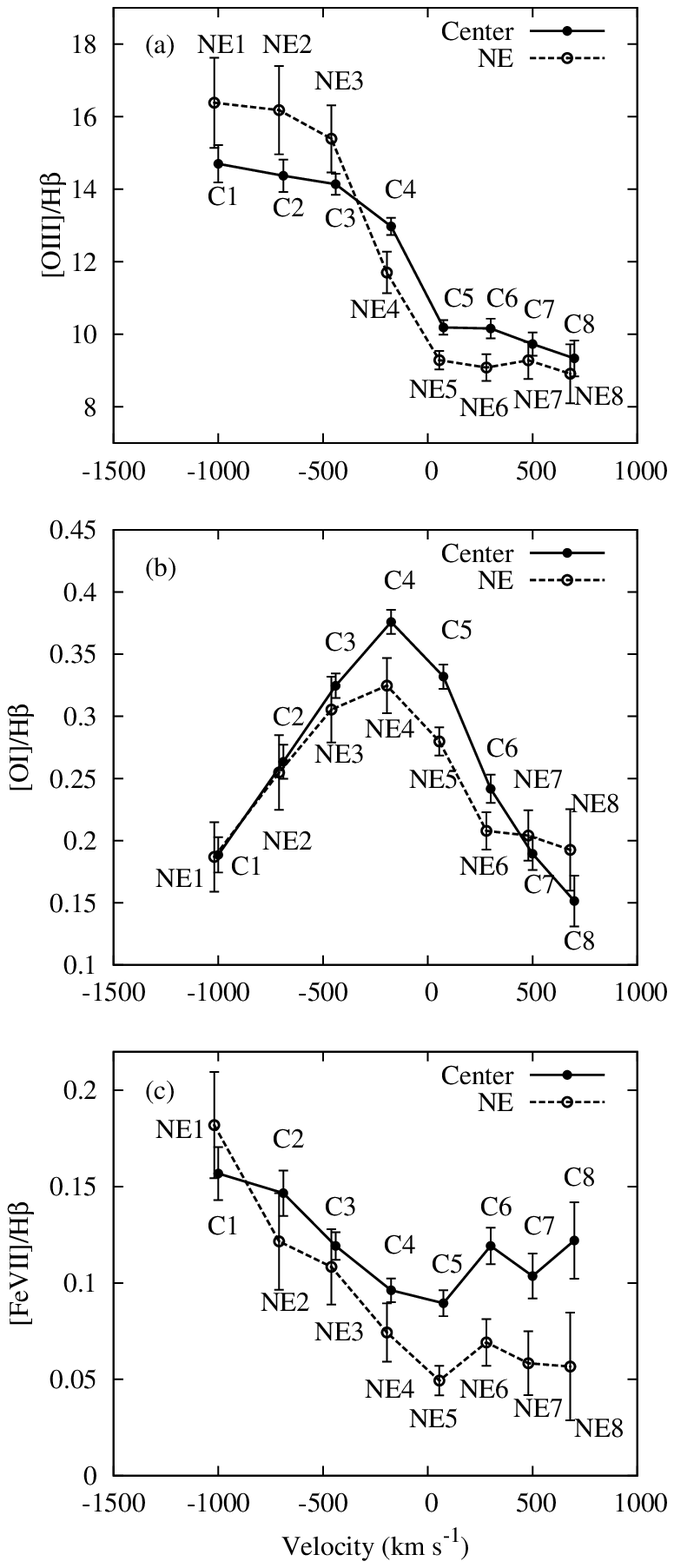}
	\end{center}
	\caption{Variation in emission-line intensity ratios with velocity. Filled circles connected by a solid line indicate the velocity bins at the center, and open circles connected by a dashed line indicate the velocity bins at the NE\timeform{3.4''}. SW components are not shown because of their low S/N ratio. Each data point is labeled with letters indicating the position and velocity bin listed in table~\ref{tab:range}.}
	\label{fig:ratio_velocity}
\end{figure}

To improve the signal-to-noise ratio, we integrated the flux over velocity bins, which were set to include the characteristic features.
Those features and the adopted velocity ranges are given in figure~\ref{fig:explain_features} and table~\ref{tab:range}.
Hereafter, the velocity bins are identified with letters indicating the location and ID number of the velocity bin, e.g., ``C1'', ``NE4''.

Figure~\ref{fig:ratio_velocity} shows the variation in emission-line intensity ratios with velocity.
Error bars indicate the one-sigma level for read-out noise and photon noise.

Uncertainties in the spectral type and velocity dispersion of the template spectra used in the stellar component subtraction were considered.
We confirmed that the uncertainties of almost all data points were smaller than, or comparable to, the error bars in figure \ref{fig:ratio_velocity} and do not affect the results significantly.
Two exceptions are C1 and NE1.
When we adopted K0III as the template spectrum, the [O\emissiontype{I}] fluxes of C1 and NE1 were approximately 1.3 times larger than the other cases.
Since the template spectra were shifted by the recession velocity of NGC~1068, those spectra included the atmospheric absorption band in the C1 and NE1 velocity range.
A stellar-origin absorption line is seen between atmospheric absorption lines in the K0III spectrum, which overlaps the atmospheric absorption lines in the other templates because of the difference in stellar radial velocities.
Hence, we may have underestimated the [O\emissiontype{I}] flux of C1 and NE1 in the adopted G5III case.
However, even if the [O\emissiontype{I}] flux of C1 and NE1 is increased to 1.3 times larger than the measured value, our discussion is not significantly affected.

To evaluate contamination of the red wing of [O\emissiontype{III}]$\lambda4959$ into the blue wing of [O\emissiontype{III}]$\lambda5007$, we shifted the [O\emissiontype{III}]$\lambda5007$ profile by a separation between two lines and scaled it down to match the [O\emissiontype{III}]$\lambda4959$ profile.
As a result, we confirmed that such contamination is negligible.

As shown by \authorcite{kc2000III} and \citet{cecil2002}, the [O\emissiontype{III}]/H$\beta$ ratio is larger in the blueshifted bins than the redshifted bins, and this feature is more remarkable at the NE\timeform{3.4''} than the center.
This was also seen in the [Fe\emissiontype{VII}]/H$\beta$ ratios.
The [O\emissiontype{I}]/H$\beta$ ratio was largest at C4 and NE4, and its value monotonically decreased with separation from C4 and NE4.

\begin{figure*}
	\begin{center}
		\FigureFile(160mm,65mm){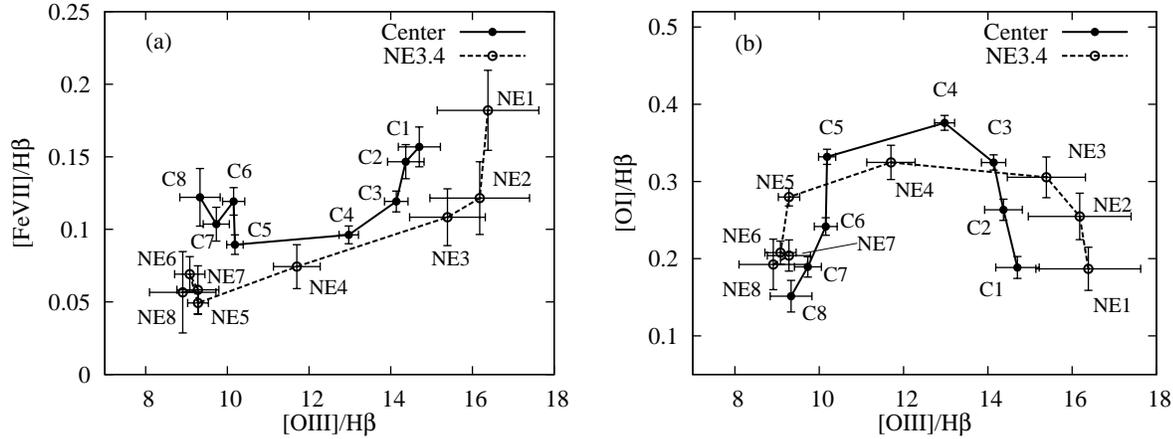}
	\end{center}
	\caption{Line ratio diagrams. (a) [O\emissiontype{III}]$\lambda 5007$/H$\beta$ vs. [Fe\emissiontype{VII}]$\lambda 6087$/H$\beta$. (b) [O\emissiontype{III}]$\lambda 5007$/H$\beta$ vs. [O\emissiontype{I}]$\lambda 6300$/H$\beta$. Filled circles connected by a solid line indicate the velocity bins at the center, and open circles connected by a dashed line indicate the velocity bins at the NE\timeform{3.4''}. Each data point is labeled with letters indicating the position and velocity bin listed in table~\ref{tab:range}.}
	\label{fig:bpt}
\end{figure*}

Figure~\ref{fig:bpt} shows line-intensity-ratio diagrams.
In figure~\ref{fig:bpt}a, the blueshifted components of the center and NE\timeform{3.4''} shift to the upper right as velocity increases with respect to the systemic velocity.
The redshifted and systemic components of the center and NE\timeform{3.4''} occupy the bottom-left region.
In figure~\ref{fig:bpt}b, the blueshifted components of both the center and NE\timeform{3.4''} shift to the lower right with increasing velocity.
The data points of the redshifted components shift lower with increasing velocity with respect to the systemic velocity.

\section{Photoionization model} \label{sec:phot_model}
In this section, we compare the observed results with a photoionization model in order to understand the physical conditions of each velocity component. Cloudy version 06.02d \citep{cloudy} was used for the model calculations.

\subsection{Assumptions and input parameters}
We adopted a power law for the ionizing continuum shape.
\citet{pier} reported that the spectral index ($\alpha$; $f_\nu \propto \nu^\alpha$) of the nuclear ionizing continuum of NGC~1068 is $-1.7$ from interpolation between the UV and the soft X-ray data, while \authorcite{kc2000III} succeeded in reproducing the observed emission-line intensity ratios with a shallower slope, $\alpha = -1.4$.
Given such uncertainties, we varied $\alpha$ in our photoionization model from $-1.7$ to $-1.4$ with an interval of 0.1.
In addition, we varied hydrogen density ($n_{\mathrm{H}}$) from $10^2$ cm$^{-3}$ to $10^6$ cm$^{-3}$ and the ionization parameter ($U$) from 10$^{-0.5}$ to 10$^{-4}$.
$U$ is given by: 
\begin{equation}
	U = \frac{Q}{4 \pi ~r^2 ~c ~n_\mathrm{H}}~,
	\label{eq:u_def}
\end{equation}
where $Q$ is the number of ionizing photons emitted per unit time, $r$ is the distance from the nucleus, and $c$ is the speed of light.
We assumed a dust-free solar chemical composition, as we could not constrain many physical parameters with only four emission lines. 
We assumed a constant-density cloud with plane-parallel geometry.
The termination criterion for the calculation was an electron temperature of 1,000 K for enough neutral region to be included; i.e., the radiation-bounded case.

\subsection{Results of the model calculation}

\begin{figure*}
	\begin{center}
		\FigureFile(160mm,125mm){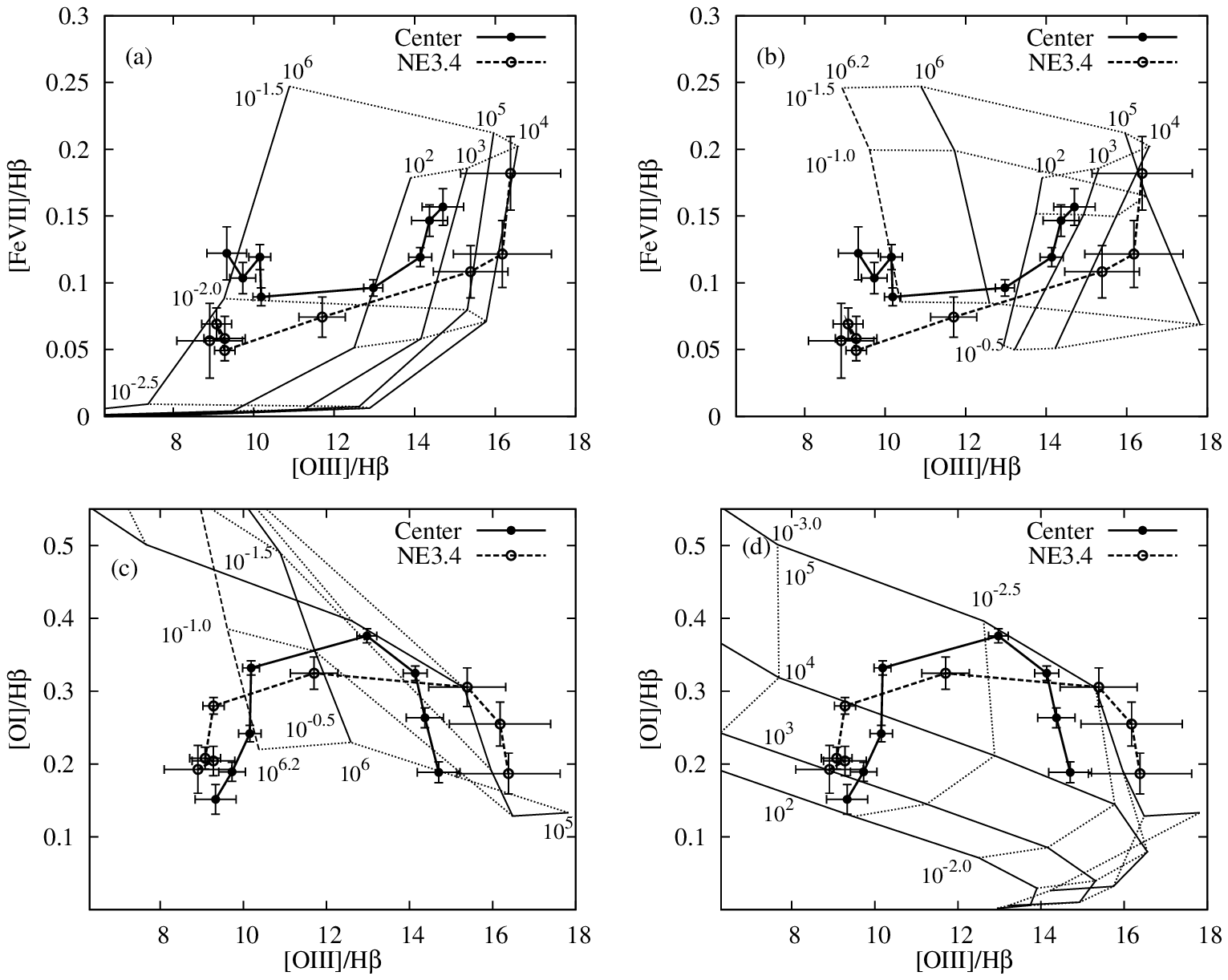}
	\end{center}
	\caption{Line ratio diagrams with the model grid.
(a)[O\emissiontype{III}]/H$\beta$ vs. [Fe\emissiontype{VII}]/H$\beta$ with $U \leq 10^{-1.5}$ and 
(b) with $U \geq 10^{-1.5}$.
(c)[O\emissiontype{III}]/H$\beta$ vs. [O\emissiontype{I}]/H$\beta$ with $n_\mathrm{H} \geq 10^5$ cm$^{-3}$ and 
(d) with $n_\mathrm{H} \leq 10^5$ cm$^{-3}$.
Dashed lines indicate results given $n_\mathrm{H}=10^{6.2}$ cm$^{-3}$.
}
	\label{fig:ratio_map}
\end{figure*}

In figure~\ref{fig:ratio_map}, the results of the model calculation for $\alpha=-1.5$ are plotted together with the observed data.
The [O\emissiontype{III}]/H$\beta$ ratio reversed at a density of about $10^5$ cm$^{-3}$ because of [O\emissiontype{III}]$\lambda 5007$ collisional de-excitation.
The [Fe\emissiontype{VII}]$\lambda6087$/H$\beta$ ratio reached a maximum at $U \sim 10^{-1.5}$ because the He$^+$ absorption suppressed the Fe\emissiontype{VII}-emitting region when $U > 10^{-1.5}$.
Owing to this feature, this ratio is a good indicator for the ionization parameter around $U=10^{-1.5}$.
The [O\emissiontype{I}]/H$\beta$ ratio increased with spectral index, since a partially ionized region becomes relatively wide compared with a fully ionized region with increasing spectral index.
The [O\emissiontype{I}]/H$\beta$ ratio decreased as the ionization parameter increased.
This is explained as follows: when the spectral index and density are fixed, the size of the partially ionized region does not depend on the number of ionizing photons because it approximates a mean free path for the ionizing photons, while the size of the fully ionized region increases with the number of ionizing photons.

[Ca\emissiontype{V}]$\lambda6087$ may overlap with [Fe\emissiontype{VII}]$\lambda6087$.
To estimate this effect, we reproduced the plot from figures~\ref{fig:ratio_map}a and \ref{fig:ratio_map}b, except that the vertical axis now represented [Fe\emissiontype{VII}]+[Ca\emissiontype{V}], and confirmed that [Ca\emissiontype{V}]$\lambda6087$ had a negligible effect on the results.

\subsection{Comparison with observations}\label{sec:ResultOfModel}

We adopted $\alpha = -1.5$ as the best-fit case.
When $\alpha=-1.7$ or $-1.6$, the [O\emissiontype{I}]/H$\beta$ ratios predicted by the model were smaller than observed.
When $\alpha=-1.5$ rather than $-1.4$, the model simultaneously reproduced more data points in the [Fe\emissiontype{VII}]/H$\beta$ vs. [O\emissiontype{III}]/H$\beta$ and [O\emissiontype{I}]/H$\beta$ vs. [O\emissiontype{III}]/H$\beta$ diagrams.
Parameters derived from model/observation comparisons are listed in table~\ref{tab:parameters}.
The comparisons are described in depth in the following subsubsections.

\begin{table*}
	\begin{center}
	\caption{Derived parameters and estimated lower limits of luminosity.}
	\label{tab:parameters}
	\begin{tabular}{cllccr@{}l} \hline
Bin	& ~~~$U$ & ~~$n_\mathrm{H}$       & Cloud & Projected & \multicolumn{2}{c}{Luminosity}\\[-4pt]
    &     & (cm$^{-3}$) & name  & distance (pc) & \multicolumn{2}{c}{($10^{45}$ erg~s$^{-1}$)}\\[2pt] \hline 
C1 & $10^{-0.9}$ & $10^{5.5}$  & F sub & 56 & ~~~~28& \\
   & $10^{-1.7}$ \footnotemark[$\dagger *$] & $10^{5.3}$ \footnotemark[$\dagger *$] &    &    & 2&.8\\
   & $10^{-1.6}$ \footnotemark[$\dagger$] & $10^{2.7}$ \footnotemark[$\dagger$] &    &    & 0&.0087\\
   & $10^{-1.1}$ \footnotemark[$\dagger$] & $10^{2.8}$ \footnotemark[$\dagger$] &    &    & 0&.035\\
C2 & $10^{-1.7}$ \footnotemark[$*$] & $10^{5.3}$ \footnotemark[$*$] & F & 56 & 2&.8 \\
   & $10^{-0.9}$ & $10^{5.5}$ &         &    & 28& \\
C3 & $10^{-1.8}$ & $10^{5.3}$ & D & 61 & 2&.6  \\
C4 & $10^{-1.9}$ & $10^{5.5}$ & E & 48 & 2&.0 \\
NE1 & $10^{-1.6}$ & $10^{5.0}$ & H & 126 & 8&.8 \\
NE2 & $10^{-1.8}$ & $10^{5.0}$ & G & 126 & 5&.6 \\
NE3 & $10^{-1.9}$ & $10^{5.0}$ & - &  -  & \multicolumn{2}{c}{-} \\[2pt]
\hline
\multicolumn{7}{@{}l@{}}{
\hbox to 0pt{\parbox{180mm}{\footnotesize
 	\rule[0pt]{0pt}{10pt}\footnotemark[$*$] Most plausible case (see text).
 	\par\noindent
 	\footnotemark[$\dagger$] The matter-bounded case (see text).
 	}\hss}}
	\end{tabular}
	\end{center}
\end{table*}

\subsubsection{The blueshifted components}
\label{sec:parameter_blue}
C3 and C4 have unique $U$ and $n_\mathrm{H}$ parameter sets.

C2 has two possibilities, $(U,~n_\mathrm{H})=(10^{-1.7},~10^{5.3})$ and $(10^{-0.9},~10^{5.5})$.
When we adopted the higher excitation case, the estimated luminosity based on those parameters was highly inconsistent with the previous result (see subsection \ref{sec:luminosity}).
Hence, the lower excitation case, $(U,~n_\mathrm{H})=(10^{-1.7},~10^{5.3})$, is more plausible.

Although C1 has a unique parameter set, ($U$, $n_\mathrm{H}$)=($10^{-0.9}$, $10^{5.5}$), the estimated luminosity was inconsistent with the previous result, similar to the higher excitation case for C2.
In the [O\emissiontype{I}] profile at the center, flux contamination from the neighboring bin to C1 could be relatively large because the feature at C1 is very weak.
If C1 is matter-bounded and does not include any [O\emissiontype{I}] flux, alternative parameter sets for this component can be derived only from the [O\emissiontype{III}]/H$\beta$ vs. [Fe\emissiontype{VII}]/H$\beta$ diagram: $(U,~n_\mathrm{H})=(10^{-1.7},~10^{5.3})$, $(10^{-1.6},~10^{2.7})$ and $(10^{-1.1},~10^{2.8})$.
Given $(U,~n_\mathrm{H})=(10^{-1.7},~10^{5.3})$, the estimated luminosity was roughly consistent with the result of \citet{pier} (see subsection \ref{sec:luminosity}), and we could easily explain the difference in central velocity among the neutral elements (see subsection~\ref{sec:marconi}).
Thus, we concluded that $(U,~n_\mathrm{H})=(10^{-1.7},~10^{5.3})$ was most plausible for C1.

The blueshifted components at the center are very dense ($10^{5.3}$ - $10^{5.5}$ cm$^{-3}$).
\citet{walsh} presented the [O\emissiontype{III}]$\lambda 4959$ and the [O\emissiontype{III}]$\lambda 5007$ profiles of the nucleus of NGC~1068 at high spectral resolution, and derived densities of $\sim 10^5$ cm$^{-3}$ for the blueshifted components, assuming an electron temperature of 15,000 K.
His estimations were slightly lower than our results.
Since the [O\emissiontype{III}] ratio depends on both density and temperature, results are affected by the assumed temperature. 
Our results were also affected by uncertainty in abundance and dust content.
In spite of these uncertainties, the difference between the derived densities is within a factor of about two.
This supports our density estimates.

The ionization parameters of the blueshifted components increase with increasing velocity with respect to the systemic velocity.
We were able to detect small variations due to the strong sensitivity of the [Fe\emissiontype{VII}]/H$\beta$ ratio to the ionization parameter.

NE1, NE2, and NE3 have slightly lower densities ($\sim 10^5$ cm$^{-3}$) than C1, C2, and C3, 
although the uncertainty is large.
These results are consistent with \citet{axon} who reported that the density of cloud G is larger than $10^{4.5}$ cm$^{-3}$ based on the [Ar\emissiontype{IV}] ratio.
NE4 does not have acceptable parameters.

\subsubsection{The redshifted and systemic components}
\label{sec:parameter_red}

For the redshifted and systemic components both at the center and NE\timeform{3.4"}, 
$U\sim10^{-0.5}$ and $n_\mathrm{H} \sim 10^{6.2}$ cm$^{-3}$ were obtained from figure~\ref{fig:ratio_map}.
However, these results are significantly inconsistent with past studies.
\citet{walsh} derived that component 6 in his paper, which corresponds to our C6, has a density of $4.6 \times 10^4$ cm$^{-3}$.
\citet{pecontal1997} reported that the narrow systemic component of the NE region corresponding to our NE5 had a density of approximately $400$ cm$^{-3}$ from the [S\emissiontype{II}]$\lambda \lambda 6717,6731$ ratio.

The reddening correction may affect the results; the reddening for the redshifted components in this study might be overcorrected, as we mentioned in subsection \ref{sec:obs_scaling}.
This overcorrection decreases the [Fe\emissiontype{VII}]/H$\beta$ and  [O\emissiontype{I}]/H$\beta$ ratios.
When we adopted a lower reddening value, the data points shifted upward in each diagram of figure~\ref{fig:ratio_map}, which meant that the estimated densities did not change significantly.
Hence, the overcorrection for reddening does not significantly affect the density estimation.

Assuming that high-density clouds with high excitation and low-density clouds with low excitation overlap, we can reproduce the observed emission-line intensity ratios of the redshifted and systemic components (see Appendix).
It should be noted that this assumption is consistent with \citet{walsh} and \citet{pecontal1997}.

\subsection{Luminosity of the central engine}
\label{sec:luminosity}
We derived the luminosity of the ionizing continuum (L) as follows:
Given a simple power-law continuum with a spectral index $\alpha$, the luminosity per unit frequency interval can be written $L_\nu = C~\nu^\alpha$, where $C$ is a constant.
$Q$ and $L$ can be expressed by the following forms:
\begin{equation}
	Q = \int_{\nu_0}^\infty \frac{L_\nu}{h~\nu} d\nu = -\frac{C}{h~\alpha}\nu_0^{\alpha}
\end{equation}
and
\begin{equation}
L = \int_{\nu_0}^\infty L_\nu ~d\nu = -\frac{C}{\alpha + 1}\nu_0^{\alpha +1} ~,
\end{equation}
where $\nu_0$ is the frequency of the hydrogen ionization limit and $h$ is Planck's constant.
From the above three equations, we find
\begin{equation} \label{eq:L}
	L = 4 \pi ~c~h~\nu_0 ~r^2 ~n_\mathrm{H}~U~\frac{\alpha}{\alpha+1} ~.
\end{equation}

Once the distance between the nucleus and the line-emitting clouds is determined, we can derive $L$ with $U$ and $n_\mathrm{H}$ estimated in subsection \ref{sec:ResultOfModel}.
\citet{cecil2002} performed medium-resolution slit-scan observations for the NGC~1068 NLR with {\it HST}, 
and presented the [O\emissiontype{III}] profiles of each NLR cloud seen in figure~\ref{fig:clouds}.
On the basis of their peak velocities and intensities, we investigated the origins of each feature in our profiles.
Although many clouds appeared in the central extraction window, 
it can be said that C1, C2, C3, and C4 are largely contributed from clouds F's sub-peak, F, D, and E, respectively.
Clouds G and H and V-shaped filament appear in the NE extracted window.
Although the profile of the V-shaped filament was not presented by \citet{cecil2002}, \citet{pecontal1997} showed that the V-shaped filament is redshifted.
NE1 and NE2 correspond to clouds H and G, respectively.
We cannot find the corresponding clouds for NE3 and NE4.
Projected distances between the nucleus and each cloud are summarized in table~\ref{tab:parameters}.
Derived lower limits of the luminosity present a wide range as shown in table~\ref{tab:parameters}.

When we adopted ($U$, $n_\mathrm{H}$)=($10^{-1.7}$, $10^{5.3}$) for C1 and C2, the range of luminosity lower limits moderated.
\citet{pier} inferred the intrinsic nuclear spectrum of NGC~1068 with UV-optical spectropolarimetric studies and X-ray studies, and estimated the ionizing continuum luminosity.
When their result is recalculated using a distance of 14.4 Mpc (they adopted 22 Mpc.), the resultant luminosity is
\begin{equation}
	L=0.36 \times 10^{45} \left( \frac{f_{\mathrm{refl}}}{0.01} \right)^{-1} \left( \frac{D}{14.4~{\rm Mpc}}\right)^2 {\rm erg~s^{-1}},
	\label{eq:pier}
\end{equation}
where the reflection fraction, $f_{\mathrm{refl}}$, is the ratio of scattered to intrinsic nuclear light, and $D$ is the distance to NGC~1068.
\citet{pier} listed the reflection fractions derived by several authors; the values range from 0.001 to 0.05.
Even if we adopt the minimum reflection fraction of 0.001, the luminosities derived from the ($U$, $n_\mathrm{H}$)=($10^{-0.9}$, $10^{5.5}$) case for C1 and C2 are one order of magnitude larger than the above estimation.
The luminosities derived from the ($U$, $n_\mathrm{H}$)=($10^{-1.7}$, $10^{5.3}$) case for C1 and C2 are consistent with \citet{pier}, assuming the minimum reflection fraction.

\section{Discussion}
\subsection{Dependence of the central velocity on the ionization potential}
\label{sec:marconi}

\begin{figure}
	\begin{center}
		\FigureFile(80mm,60mm){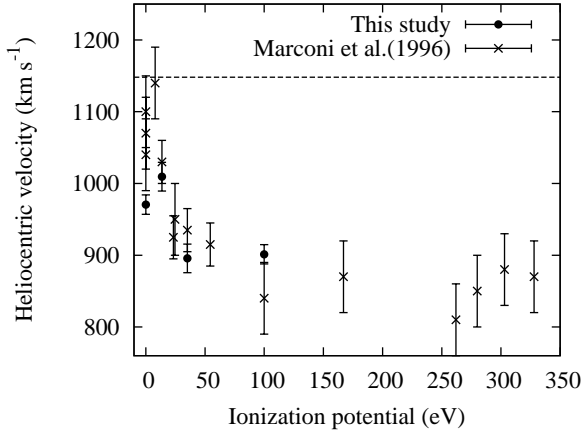}
	\end{center}
	\caption{Dependence of the heliocentric velocity of the emission lines on the ionization potential. The filled circles and crosses represent the measurements of this study and \citet{marconi}, respectively, and the dashed line marks the systemic velocity of NGC~1068.}
	\label{fig:ipvel}
\end{figure}

\begin{table}
	\begin{center}
	\caption{Neutral line critical densities}
	\label{tab:ncrit}
		\begin{tabular}{cc} \hline
		\rule[0pt]{0pt}{10pt}Line name & Critical density (cm$^{-3}$)\\[2pt] \hline
		\rule[0pt]{0pt}{12pt}~[O\emissiontype{I}]$\lambda$6300 & $1.5 \times 10^6$ \\
		~[N\emissiontype{I}]$\lambda$5200 & $7.0 \times 10^2$ \\
		~[N\emissiontype{I}]$\lambda$5198 & $2.2 \times 10^3$ \\
		~[C\emissiontype{I}]$\lambda$9850 & $1.4 \times 10^4$ \\[2pt] \hline
		\end{tabular}
	\end{center}
\end{table}

We verified the dependence of the central velocity of the emission lines on the ionization potential shown by \citet{marconi}. 
They showed that the higher excitation lines of the NGC~1068 nucleus are more strongly blueshifted with respect to the systemic velocity, and that the low-excitation lines are located close to the systemic velocity.
This correlation was also reported by \citet{kc2000II} and \citet{lutz}.
Figure \ref{fig:ipvel} shows the correlation between the Gaussian centroid velocity and the ionization potential in this study together with the results of \citet{marconi}.
Our data, except [O\emissiontype{I}], agree with the results of \citet{marconi} within the error limits.

However, we found a discrepancy in the neutral elements.
\citet{marconi} used [N\emissiontype{I}]$\lambda5200$ and [C\emissiontype{I}]$\lambda9850$ for the neutral elements, whereas the neutral element in our data was [O\emissiontype{I}].
We calculated the critical electron densities for these neutral lines with the {\em ionic} task in IRAF, assuming $T_\mathrm{e} = 10^4$ K.
The calculated densities are tabulated in table \ref{tab:ncrit}.
The critical electron density of [O\emissiontype{I}]$\lambda6300$ is about two orders of magnitude larger than the others.
Therefore, collisional de-excitation may suppress the [N\emissiontype{I}] and [C\emissiontype{I}] flux, leading to an [O\emissiontype{I}] velocity that is bluer by comparison with [N\emissiontype{I}] and [C\emissiontype{I}].
This interpretation is consistent with the result that, when we adopt ($U$, $n_\mathrm{H}$)=($10^{-1.7}$, $10^{5.3}$) for C1, the blueshifted components at the center have densities higher than $10^5$ cm$^{-3}$. 

\subsection{ NLR kinematic and excitation structure}
In the previous section, we found that $U$ of the blueshifted components increases with increasing velocity.
Given radiative acceleration, we expected to find that $U$ is velocity-dependent.
In the radiation-bounded case adopted in our photoionization model, the radiative acceleration can be estimated by
\begin{equation}\label{eq:ar}
a_{r} = \frac{L}{4 ~\pi ~r^2 ~c ~N ~m_\mathrm{p}}~,
\end{equation}
where $N$ is the column density of the cloud and $m_p$ is the proton mass.
From equations (\ref{eq:L}) and (\ref{eq:ar}), we find
\begin{equation}\label{eq:ar2}
a_{r} = \frac{h ~\nu_0}{m_\mathrm{p}} ~\frac{\alpha}{\alpha+1} \frac{U}{l}~,
\end{equation}
where $l$ is the cloud depth. 
This means that, when an optically thick cloud is accelerated by radiation from the nucleus, its acceleration is proportional to the ionization parameter and inversely proportional to cloud depth.
We assumed the radiation pressure was the driving force of the outflow in the following sections, and investigated the kinematic and excitation structure of the NLR of NGC~1068.

\subsubsection{Ionizing photon attenuation}
\label{sec:attenuation}
The velocity dependence of $U$ is expected when absorbing matter with varying column densities exists between the nucleus and the NLR clouds.
In this situation, the clouds irradiated by the more attenuated ionizing continuum are expected to have lower $U$ and lower velocity.

\begin{figure}
	\begin{center}
		\FigureFile(80mm,185mm){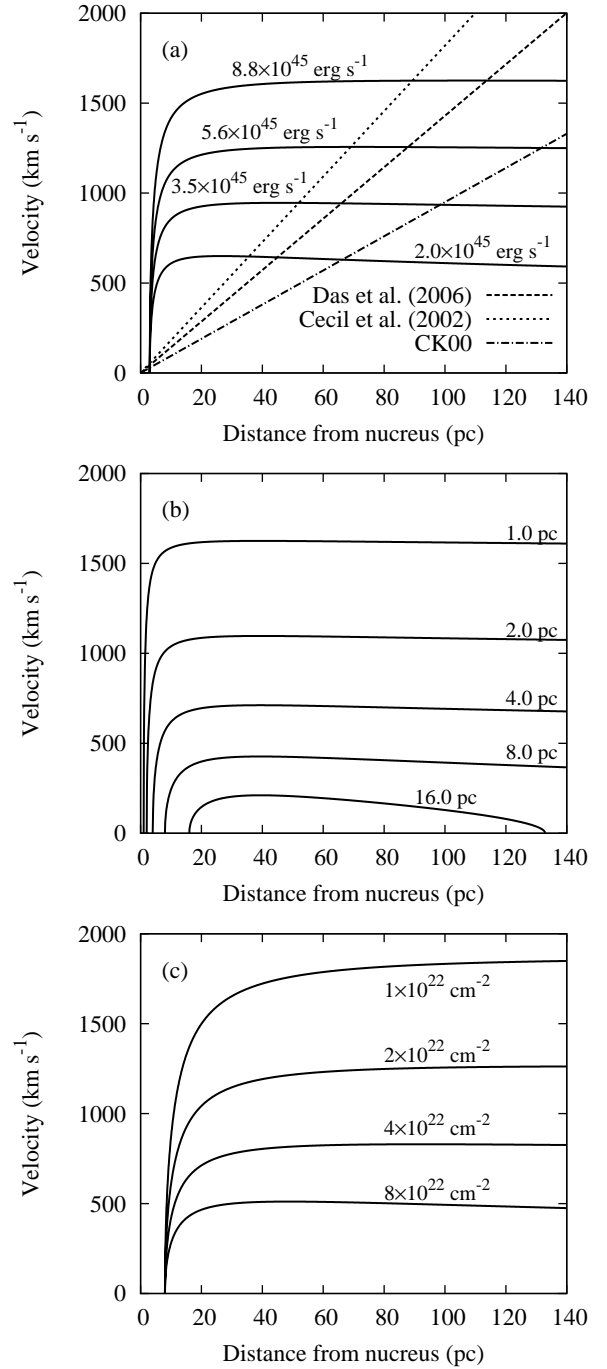}
	\end{center}
	\caption{The estimated velocities. a) Luminosity is varied with a fixed launching radius (3 pc) and column density ($10^{23}$ cm$^{-3}$). Deprojected radial dependences of the velocity reported by \citet{das2006}, \citet{cecil2002} and \authorcite{ck_kinematics2000} are also shown. b) Launching radius is varied with a fixed luminosity ($3 \times 10^{45}$ erg~s$^{-1}$) and column density ($10^{23}$ cm$^{-3}$). c) Column density is varied with a fixed luminosity ($3 \times 10^{45}$ erg~s$^{-1}$) and launching radius (8 pc).}
	\label{fig:velocity}
\end{figure}

\begin{figure*}
	\begin{center}
		\FigureFile(140mm,7.3mm){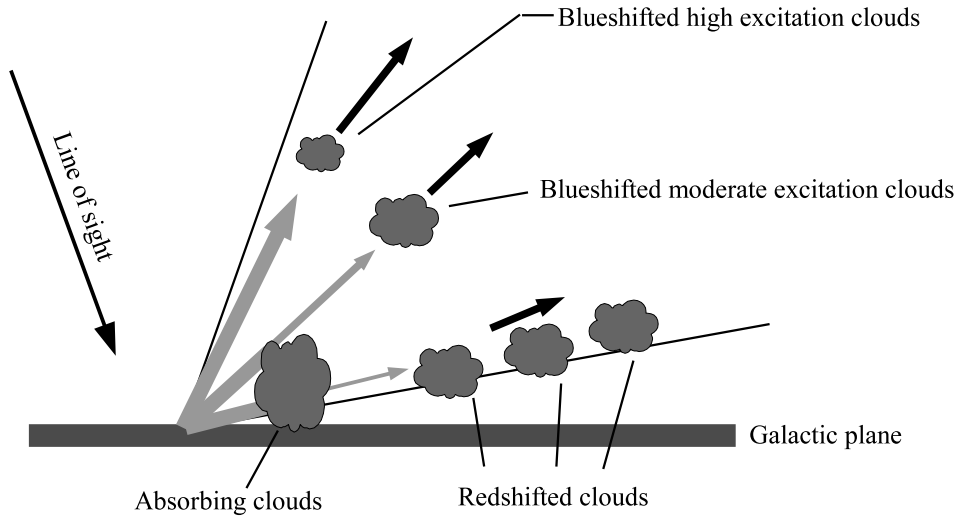}
	\end{center}
	\caption{Inferred NLR structure of NGC~1068 from the variable attenuation model together with the hollowed biconical outflow model. Black and gray arrows show velocities and ionizing radiation whose width means flux, respectively.}
	\label{fig:ponti1}
\end{figure*}

NGC~1068 probably has ionizing continuum absorbers in the vicinity of its nucleus.
\citet{alexander} and \citet{spinoglio} argued that the observed infrared emission-line ratios were reproduced more accurately by an attenuated ionizing continuum than by other continuum shapes.
From optical and UV spectroscopic studies, \authorcite{kc2000III} reported that the redshifted components are ionized by the attenuated continuum, though the blueshifted components are irradiated directly.
Their inferred absorbers for the redshifted components of NGC~1068 
attenuated the ionizing continuum by about one order of magnitude.
Hence, variation in absorbing column density seems to be able to account for the difference in $U$ observed in NGC~1068.

To allow for cloud outflow, radiative acceleration must overcome gravity.
We used the same enclosed central region mass distribution for NGC~1068 as \citet{das2007}; it is given by:
\begin{eqnarray}\label{eq:mr}
	M(r) &= & 1.5 \times 10^7 + 7.1 \times 10^6 ~r \nonumber\\	
 && \qquad + 3.2 \times 10^{10} \left( \frac{r}{r+2400} \right) ^{1.5} ~M_{\solar}~,
\end{eqnarray}
where $r$ is distance from the nucleus in pc.
The first term in equation (\ref{eq:mr}) represents the contribution from the supermassive black hole, the second term from the bulge, and the last term from the central star cluster.

When the cloud is assumed to be launched with a velocity of 0 km~s$^{-1}$ at a distance of $r_0$,
we can find the velocity at a distance $r$ from equations (\ref{eq:ar}) and (\ref{eq:mr}): 
\begin{equation}
	v(r) = \sqrt{f_1(r) + f_2(r) + f_3(r)}~,
\end{equation}
where
\begin{eqnarray}
	f_1(r) &=& \left( 10.3 ~\frac{L_{45}}{N_{23}} - 1.3 \right) \times 10^5 \left( \frac{1}{r_0} - \frac{1}{r} \right)~, \\
	f_2(r) &=& - 6.1 \times 10^4 ~\log \frac{r}{r_0}~, \\
	f_3(r) &=& - 2.2 \times 10^5 \left( \sqrt{\frac{r}{r+2400}} - \sqrt{\frac{r_0}{r_0+2400}} \right) ~,
\end{eqnarray}
where $r_0$ is in units of pc, $L_{45}$ is the ionizing continuum luminosity in units of $10^{45}$ erg~s$^{-1}$, and $N_{23}$ is the column density of the cloud in units of $10^{23} $ cm$^{-2}$.
When $n_\mathrm{H}=10^5$ cm$^{-3}$ and $U=10^{-1.5}$, the column density of the calculated cloud is about $10^{23} $ cm$^{-2}$.
However, this value is affected by the termination criterion of the calculation.
If the cloud has more neutral gas, the velocity decreases.
It should be noted that, if the column density is more than $10^{22}$ cm$^{-2}$, the [O\emissiontype{I}]-emitting zone is sufficiently included for $n_\mathrm{H}=10^5$ cm$^{-3}$ and $U=10^{-1.5}$.

We show the estimated velocities given various luminosities estimated in subsection \ref{sec:luminosity} in figure \ref{fig:velocity}a.
The estimated velocity range is $\sim$300 km~s$^{-1}$ for luminosities derived for the blueshifted components at the center, and is smaller than the observed ($\sim$ 700 km~s$^{-1}$).
This discrepancy indicates that the observed velocity range cannot be reproduced by this simple model.

\subsubsection{Projection effect}

\begin{figure*}
	\begin{center}
		\FigureFile(106.8mm,61mm){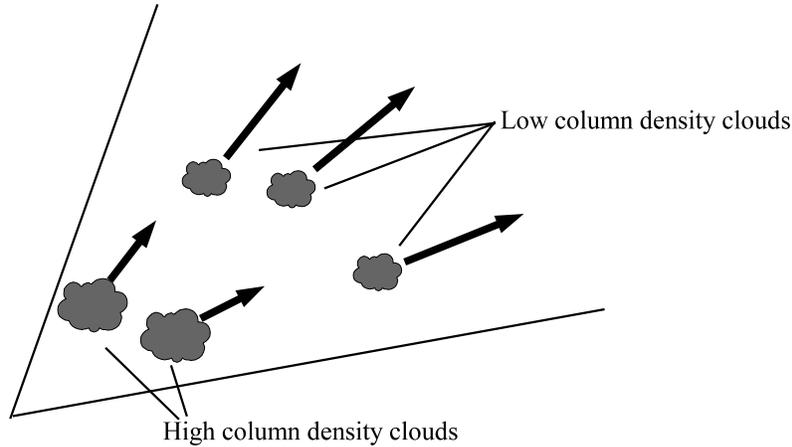}
	\end{center}
	\caption{Inferred NLR structure of NGC~1068 based on the variable column density model. Black arrows show velocities.}
	\label{fig:ponti2}
\end{figure*}

Next, we considered projection effects as well as various attenuations of the ionizing continuum.
We assumed the three-dimensional structure of the NLR in order to resolve the projection effect.
A hollowed biconical geometry was demonstrated for the NGC~1068 NLR (\cite{das2006,cecil2002}; \authorcite{ck_kinematics2000}). 
According to that geometry, the cone axis is almost perpendicular to the line of sight, and 
the farthest edge of the cone is approximately aligned with the galactic disk (see figure \ref{fig:ponti1}).

The hollowed biconical geometry explains the observed velocity difference.
When the clouds move with nearly the same velocity, cloud E, located close to the limb of the ionized cone, has a lower velocity than clouds 
D and F, located close to the center of the ionized cone.
In the hollowed biconical geometry, the blueshifted clouds with lower velocity are located nearer to the galactic disk than those with higher velocity.
If the absorbers lie in the galactic disk, 
the absorbing column density decreases outward from the galactic disk,
resulting in the observed velocity dependence of $U$.

Figure \ref{fig:velocity}a also shows deprojected radial dependence of velocities reported by \citet{das2006}, \citet{cecil2002} and \authorcite{ck_kinematics2000}.
We see that, even if we vary the luminosity, launching radius, or column density parameters, the gradual increase in velocity with radius cannot be reproduced (figure \ref{fig:velocity}), which is also reported by \citet{das2007}.
We have to include another perspective to explain the velocity field.

\subsubsection{New perspective}

\begin{figure}
	\begin{center}
		\FigureFile(80mm,185mm){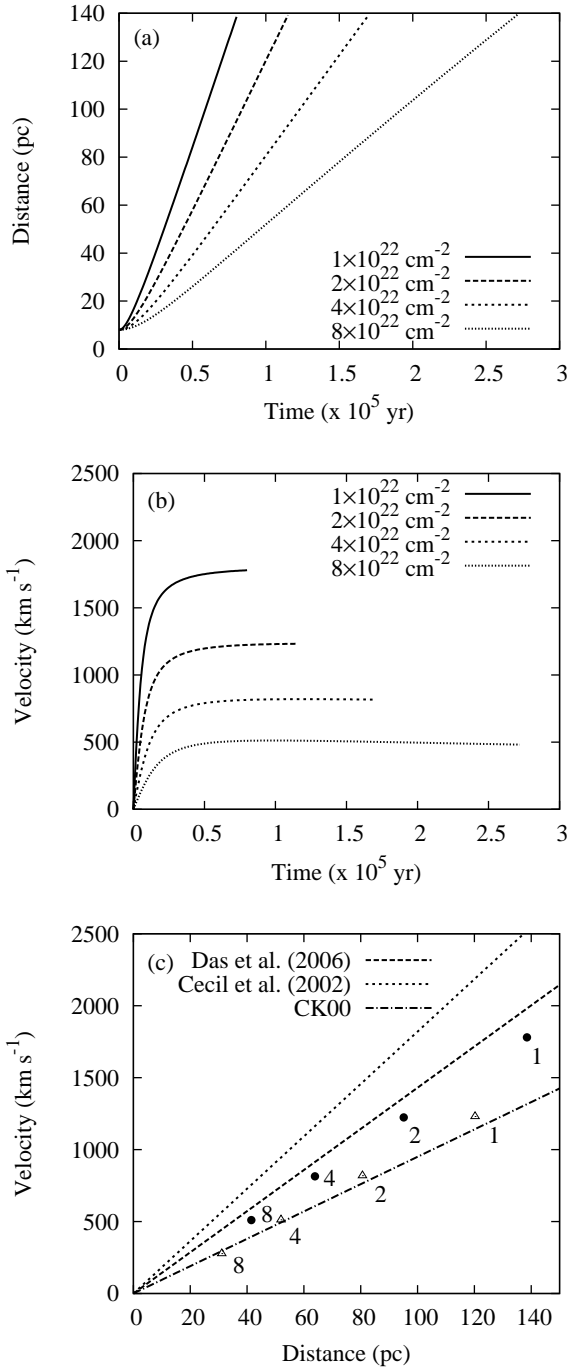}
	\end{center}
	\caption{Time dependence of distance (a) and velocity (b) of the clouds with various column densities, when $L_{45}=3.0$ and $r_0 = 8$ pc. In the bottom panel (c), solid circles show estimated distances and velocities of clouds with various column densities after $8 \times 10^4$ years since launching. Labels with the circles show the column density in units of $10^{22}$ cm$^{-2}$. 
Open triangles show the results in the half luminosity case in which $r_0 = 8$ pc and time interval since launching is $1 \times 10^5$ years.
The radial dependence of the velocity reported by \citet{das2006}, \citet{cecil2002} and \authorcite{ck_kinematics2000} is also shown.}
	\label{fig:time_dist_vel_col}
\end{figure}

We propose a new perspective to explain the radial dependence of the velocity.
Up to now, we have considered the observed velocity field as a time-sequence of cloud motion.
When clouds with various column densities are launched simultaneously at the same radius, clouds with lower column densities attain higher velocities and are located farther from the center for a fixed time (figure \ref{fig:ponti2}).
This situation might give rise to the observed velocity field.

We estimated the velocity field inferred through such a situation, and showed the results in figure \ref{fig:time_dist_vel_col}.
In this model, we adopted the typical luminosity ($3 \times 10^{45}$ erg s$^{-1}$) estimated in section \ref{sec:luminosity}.
Figures \ref{fig:time_dist_vel_col}a and b show the time-dependence of distance and velocity, respectively.
Figure \ref{fig:time_dist_vel_col}c shows the distances and velocities of clouds with various column densities $8 \times 10^4$ years after launching.
It is impressive that the estimations match the previously reported velocity field well.

There are uncertainties of the luminosity estimations in section \ref{sec:luminosity} because of our data qualities and uncertainties of our photoionization model.
Those luminosity uncertainties affect the results in figure \ref{fig:time_dist_vel_col}.
For different luminosities, however, we can obtain similar results with different launching radii or different time intervals from launching.
As an example, we also show the results in the half luminosity case in figure \ref{fig:time_dist_vel_col}c.
It is noted that, in extremely low luminosity case, the clouds cannot outflow.

The above situation could occur when a molecular cloud with many cloud cores of various column densities encounters the ionization cone.
\citet{orion-a} observed the Orion-A molecular cloud with the H$^{13}$CO$^+$({\it J}=1-0) molecular line, which has a high critical density ($8 \times 10^4$ cm$^{-3}$).
They identified 236 dense cores, and presented a mean core radius of 0.14$\pm$0.03 pc and a mean density of (1.6$\pm$1.2)$\times 10^4$ cm$^{-3}$.
From these values, we estimated the core column density as $7 \times 10^{21}$ cm$^{-2}$.
This estimated column density is roughly consistent with the adopted values in figure~\ref{fig:time_dist_vel_col}.

The maximum core diameter in \citet{orion-a} is 0.46 pc, but the cloud sizes are measured as about 10 pc in figure \ref{fig:clouds}.
This might suggest that the NLR clouds are gatherings of small cloudlets.
This idea is supported by the filling factor of the NLR clouds.
\citet{cecil2002} listed the areas, the [O\emissiontype{III}]$\lambda 5007$ luminosities and the [O\emissiontype{III}]$\lambda 5007$/H$\beta$ ratios for each NLR cloud.
We estimated the filling factor to be very small ($\sim 10^{-5}$) from those values and the densities derived in section \ref{sec:phot_model}.
In this estimation, we adopted $1.24 \times 10^{-25}$ erg~s$^{-1}$~cm$^{3}$ as $4 \pi j_{H\beta}/n_e n_p$ where $j_{H\beta}$ is the H$\beta$ emission coefficient, and $n_\mathrm{e}$ and $n_\mathrm{p}$ are electron and proton densities, respectively \citep{agnagn}.
This is the value when Case B and an electron temperature of 10,000 K are assumed.
We assumed $n_\mathrm{e} = n_\mathrm{p} = n_\mathrm{H}$.

The timescale during which an outflowing cloud with 500 km~s$^{-1}$ passes through the NLR ($\sim$100 pc) is about $10^5$ years. This is less than the inferred AGN lifetime of $10^6$ - $10^8$ years (\cite{lifetime} and references therein).
Hence, if the above situation is accurate, we are observing a transient phenomenon.
The narrow lines of NGC~1068 are abnormally broader than the other Seyfert galaxies.
This abnormality can be explained by the transient phenomenon of dense cloud outflow.

In conclusion, we can reproduce both the velocity dependence of $U$ and the velocity field of the NLR of NGC~1068 through varying ionizing continuum attenuations, a hollowed biconical geometry, and this new perspective.

\section{Summary}

We performed medium-resolution spectroscopic observations of NGC~1068 and obtained 
emission-line profiles of [O\emissiontype{III}]$\lambda 5007$, H$\beta$, [Fe\emissiontype{VII}]$\lambda 6087$ and [O\emissiontype{I}]$\lambda 6300$.
At the center, [O\emissiontype{III}], H$\beta$ and [O\emissiontype{I}] exhibited a peak at $-170$ km~s$^{-1}$ with respect to the systemic velocity.
The double-peaked profile of [Fe\emissiontype{VII}] was confirmed as reported by \citet{rodriguez}. Its two peaks at the center were located at the sub-peaks of [O\emissiontype{III}] and H$\beta$.
We marginally detected the narrow spike component in the [Fe\emissiontype{VII}] profile at the NE\timeform{3.4''}.
In the [O\emissiontype{I}] profile at the center, the features of the high-velocity components were very weak.

Comparing observations with a photoionization model, we investigated the physical conditions of the line-emitting regions.
The ionization parameters of the blueshifted components increased with increasing velocity with respect to the systemic velocity.
The bluest components at the center might be matter-bounded.
The densities of the blueshifted components at the center were $\sim 10^{5.3}$ - 10$^{5.5}$ cm$^{-3}$ and those at the NE\timeform{3.4''} were slightly lower than at the center.
The systemic and redshifted components might be constructed from two or more regions with different excitation states.

When we assumed a hollowed biconical geometry in which absorbing matter with varying column densities appears between the nucleus and the NLR clouds, we succeeded in reproducing the velocity dependence of the ionization parameter.
However, this model did not explain the gradual increase in the velocity field with radius.
We succeeded in reproducing the velocity field given clouds with various column densities launched simultaneously at the same radius.
Finally, we show that both the observed velocity dependence of the ionization parameter and the gradually increasing velocity field can be reproduced by varying the ionizing continuum attenuation, assuming a hollowed biconical geometry and varying the column densities of outflowing clouds.
\bigskip

The author would like to thank M.~Yoshida, I.~Iwata, H.~Sugai, H.~Ohtani, T.~Ishigaki, T.~Kawaguchi and Kyoto 3DII group for useful discussions.
The author wishes to thank K.~Tanaka, A.~Arai, M.~Katsuura, and M.~Kamata for support during the observations at NHAO.
The author also thank Gerald Cecil for his helpful comments as the referee.
The observations were carried out when the author belonged to NHAO.

Some of the data presented in this paper were obtained from the Multimission Archive at the Space Telescope Science Institute (MAST). STScI is operated by the Association of Universities for Research in Astronomy, Inc., under NASA contract NAS5-26555. Support for MAST for non-{\it HST} data is provided by the NASA Office of Space Science via grant NAG5-7584 and by other grants and contracts.

This paper makes use of data obtained from the Isaac Newton Group Archive which is maintained as part of the CASU Astronomical Data Centre at the Institute of Astronomy, Cambridge.

\appendix
\section*{Clouds located along a line-of-sight}

\begin{figure*}
	\begin{center}
		\FigureFile(160mm,60mm){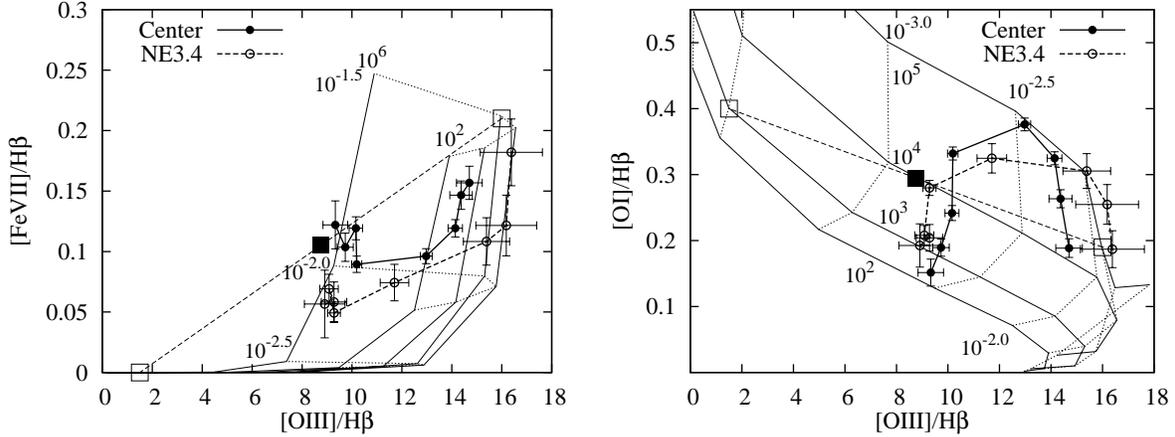}
	\end{center}
	\caption{Line ratio diagrams. Model grid and observed data points are the same as figures \ref{fig:ratio_map}a and d.
Open and filled squares indicate, respectively, the locations of each cloud and data point that may be observed.}
	\label{fig:mix}
\end{figure*}

In this appendix, we show the expected location of observed data in an emission-line intensity ratio diagram given clouds with different physical conditions located along a line-of-sight.
For simplicity, we consider two clouds, clouds 1 and 2.
We write fluxes of H$\beta$ and [O\emissiontype{III}] as $F$(H$\beta$) and $F$([O\emissiontype{III}]), respectively.
The total fluxes are given by
\begin{equation}
	F\mathrm{_{tot}(H\beta)} = F\mathrm{_1(H\beta)} + F\mathrm{_2(H\beta)}
\end{equation}
and
\begin{equation}
	F\mathrm{_{tot}([O\emissiontype{III}])}=F\mathrm{_1([O\emissiontype{III}])} + F\mathrm{_2([O\emissiontype{III}])} ~.
\end{equation}
Assuming the ratio of the H$\beta$ flux of cloud 1 to total H$\beta$ flux is $\eta$, we get
\begin{equation}
	\frac{F\mathrm{_1(H\beta)}}{F\mathrm{_{tot}(H\beta)}} = \eta ~~and~~ \frac{F\mathrm{_2(H\beta)}}{F\mathrm{_{tot}(H\beta)}} = 1 - \eta ~.
\end{equation}
From the above equations, we find
\begin{equation}
	\frac{F\mathrm{_{tot}([O\emissiontype{III}])}}{F\mathrm{_{tot}(H\beta)}} = \eta \frac{F\mathrm{_1([O\emissiontype{III}])}}{F\mathrm{_1(H\beta)}} + (1-\eta)\frac{F\mathrm{_2([O\emissiontype{III}])}}{F\mathrm{_2(H\beta)}}~.
\end{equation}
This equation is also applicable to other emission-line intensity ratios, [Fe\emissiontype{VII}]/H$\beta$ and [O\emissiontype{I}]/H$\beta$.
This result indicates that, when we connect the locations of each cloud in the diagram with a line, the observed data are located at a point dividing the line into $\eta$:$(1-\eta)$.

For example, when ($U$, $n_{\mathrm{H}}$)=($10^{-1.5}$, $10^{5}$) for cloud 1 and ($10^{-3.5}$, $10^{3}$) for cloud 2, and when those clouds contribute equally to the total H$\beta$ flux, we expect that the observed locations are the black squares in figure \ref{fig:mix}.

\end{document}